\documentstyle[12pt,epsf]{article}
\newlength{\dinwidth} \newlength{\dinmargin}
\begin{document}
\begin{titlepage}
  \begin{flushright}
    \begin{tabular}[h]{l}
      FTUAM-93/01 \\
      CRN/PT 93-39\\
      nucl-th/9307001\\
      October, 1993
    \end{tabular}
\vspace{1cm}
  \end{flushright}
  \begin{center}
    \LARGE{A Full $pf$ Shell Model Study of A~=~48
      Nuclei}\\[0.7cm]
    {\large {\bf E. Caurier, A.\,P. Zuker}\\[0.2cm]
    Groupe de Physique Th\'eorique\\
      CRN IN2P3-CNRS/Universit\'e Louis Pasteur BP20\\
      F-67037 Strasbourg-Cedex, France \\[0.5cm]
      {\bf A. Poves and G. Mart\'{\i}nez-Pinedo}\\[0.2cm]
      Departamento de F\'{\i}sica Te\'orica C-XI\\
      Universidad Aut\'onoma de Madrid\\
      E-28049 Madrid, Spain.\\}
  \end{center}

\vspace{1cm}

\begin{abstract}
Exact diagonalizations with a minimally modified realistic
force lead to detailed agreement with measured level schemes and
electromagnetic transitions in $^{48}$Ca, $^{48}$Sc, $^{48}$Ti,
$^{48}$V, $^{48}$Cr and $^{48}$Mn. Gamow-Teller strength functions are
systematically calculated and reproduce the data to within the
standard quenching factor. Their fine structure indicates that
fragmentation makes much strength unobservable. As a by-product, the
calculations suggest a microscopic description of the onset of
rotational motion. The spectroscopic quality of the results provides
strong arguments in favour of the general validity of monopole
corrected realistic forces, which is discussed.

\end{abstract}

\end{titlepage}

\section{Introduction}

Exact diagonalizations in a full major oscillator shell are the
privileged tools for spectroscopic studies up to $A\approx60$.
The total number of states ---$2^d$ with $d=12$, 24 and 40 in the $p$,
$sd$, and $pf$ shells respectively--- increases so fast that three
generations of computers and computer codes have been necessary to
move from $n=4$ to $n=8$ in the $pf$ shell i.e.\ from four to eight
valence particles, which is our subject.

A peculiarity of the $pf$ shell is that a minimally modified realistic
interaction has been waiting for some 15 years to be tested in
exact calculations {\it with sufficiently large number of particles},
and $n=8$ happens to be the smallest for which success was practically
guaranteed. In the test, spectra and electromagnetic transitions will
be given due place but the emphasis will go to processes governed
by spin operators: beta decays, $(p,n)$ and $(n,p)$ reactions.
They are interesting ---perhaps fascinating is a better word---
on two counts. They demand a firm understanding of not simply a few,
but very many levels of given $J$ and they raise the problem of
quenching of the Gamow-Teller strength.

As by-products, the calculations provide clues on rotational motion
and some helpful indications about possible truncations of the
spaces. The paper is arranged as follows.

Section 2 contains the definition of the operators. Some preliminary
comments on the interaction are made.

In each of the following six sections, next to the name of the
nucleus to which it is devoted, the title contains a comment
directing attention to a point of interest. The one for $^{48}$Ca is
somewhat anomalous.

In section 9, the evidence collected on GT strength is analyzed. Our
calculations reproduce the data once the standard quenching factor,
$(0.77)^2$, is adopted. The fine structure of the strength function
indicates that fragmentation makes impossible the observation of many
peaks. Several experimental checks are suggested. Minor discrepancies
with the data are attributed to small uncertainties in the $\sigma
\cdot \sigma$ and $\sigma \tau \cdot \sigma\tau$ contributions to the
force.

In section 10 we examine the following question:

\noindent
{\it Why, in the sd shell, phenomenologically fitted matrix elements
have been so far necessary to yield results of a quality comparable
with the ones we obtain here with a minimally modified realistic
interaction?}
The short answer is that {\it monopole} corrected realistic forces are
valid in general, but the fact is easier to detect in the $pf$ shell.

Section 11 contains a brief note on binding energies. In section 12 we
conclude.

The rest of the introduction is devoted to a point of notation, a
review of previous work (perhaps not exhaustive enough, for which we
apologize) and a word on the diagonalizations.

\noindent
{\bf Notations.} Throughout the paper $f$ stands for
$f_{7/2}$ (except of course when we speak of the $pf$ shell)
and $r$, generically, for any or all of the other subshells
($p_{1/2}\;p_{3/2}\;f_{5/2}$). Spaces of the type

\begin{equation}
f^{n-n_0} r^{n_0}+f^{n-n_0-1} r^{n_0+1}+\cdots+
f^{n-n_0-t} r^{n_0+t}
\end{equation}
represent possible truncations: $n_0$ is different from zero if more
than 8 neutrons are present and when $t=n-n_0$ we have the full space
$(pf)^n$ for A~=~$40+n$.

\noindent
{\bf Bibliographical note.} The characteristic that makes
the $pf$ shell unique in the periodic table is that at $t=0$ we
already obtain a very reasonable model space, as demonstrated in the
$f^n$ case (i.e. $n_0=0$) by Ginocchio and French~\cite{gifr} and
Mc~Cullen, Bayman and Zamick~\cite{mbz} (MBZ in what follows). The
$n_0\not=0$ nuclei are technically more demanding but the $t=0$
approximation is again excellent (Horie and Ogawa~\cite{hoog}).

The first systematic study of the truncation hierarchy was undertaken
by Pasquini and Zuker~\cite{pazu1,pazu2} who found that $t=1$ has
beneficial effects, $t=2$ may be dangerous and even nonsensical, while
$t=3$ restored sense in the only non trivial case tractable at the
time ($^{56}$Ni).  This seems to be generally a fair approximation,
and the topic will be discussed as we proceed.

Much work on $n_0 \neq 0$ nuclei with $t=1$ and 2 spaces was done by
the groups in Utrecht (Glaudemans, Van Hees and
Mooy~\cite{glau1,glau2,glau3} and Tokyo (Horie, Muto,
Yokohama~\cite{hori1,hori2,hori3}).

For the Ca isotopes high $t$ and even full space diagonalizations have
been possible (Halbert, Wildenthal and Mc~Grory~\cite{halb1,halb2}).

Exact calculations for $n\leq 4$ are due to Mc~Grory~\cite{mcgr}, and
some $n=5$ nuclei were studied by Cole~\cite{cole} and by Richter, Van
der Merwe, Julies and Brown in a paper in which two sets of interaction
matrix elements are constructed~\cite{rich}.

For $n=6,\,7$ and 8 the only exact results reported so far are those
of the authors on the magnetic properties of the Ti
isotopes~\cite{cpz1}, and the double $\beta$ decay calculations of
$^{48}$Ca by Ogawa and Horie~\cite{ogho}, the authors~\cite{cpz2} and
Engel, Haxton and Vogel~\cite{enge}. The MBZ model has been reviewed
in an appendix to Pasquini's thesis~\cite{pazu1} and by Kutschera,
Brown and Ogawa~\cite{kubro}.  Its success suggested the
implementation of a perturbative treatment in the full $pf$ shell by
Poves and Zuker~\cite{pozu1,pozu2}.

{\it All the experimental results
for which no explicit credit is given come from the recent compilation
of Burrows for} A~=~48~\cite{burrows}.

\begin{table}[h]
  \begin{center}
    \leavevmode
    \begin{tabular}{lccccc}
      \hline \hline
      & $^{48}$Ca & $^{48}$Sc & $^{48}$Ti & $^{48}$V & $^{48}$Cr \\
      \hline
      m-scheme & 12\,022 & 139\,046 & 634\,744 & 1\,489\,168 &
      1\,963\,461 \\
      \hline
      J T & 4 4 & 4 3 & 4 2 & 5 1& 4 0 \\
      & 1\,755 & 17\,166 & 63\,757 & 106\,225 & 58\,219\\
      \hline \hline
    \end{tabular}
  \end{center}
  \caption{m-scheme and maximal $JT$ dimensions in the full $pf$ shell}
  \label{tab:dimensiones}
\end{table}

\noindent {\bf The diagonalizations} are performed in the $m$-scheme
using a fast implementation of the Lanczos algorithm through the code
ANTOINE~\cite{antoine}. Some details may be found in ref.~\cite{cpz3}.
The strength functions are obtained through Whitehead's
prescription~\cite{white}, explained and illustrated in
refs.~\cite{cpz1,cpz2} (and section 9).  The $m$-scheme and maximal
$JT$ dimensions of the nuclei analyzed are given in table~1.  To the
best of our knowledge they are the largest attained so far.

\section{The Interaction and other Operators}

The most interesting result of refs.~\cite{pazu1,pazu2} was that the
spectroscopic catastrophes generated by the Kuo-Brown
interaction~\cite{kuob} in some nuclei, could be cured by the simple
modification (KB' in~\cite{pazu1,pazu2}).

\begin{equation}
V_{fr}^T{\mbox{(KB1)}}=V_{fr}^T{\mbox{(KB)}}-(-)^T\,300\,\mbox{keV}
\end{equation}
where $V_{fr}^T$ are the centroids, defined for any two shells by

\begin{equation}
V_{ij}^T=\frac{\sum_J(2J+1)W_{ijij}^{JT}}{\sum_J(2J+1)}\,,
\end{equation}
where the sums run over Pauli allowed values if $i=j$, and
$W_{ijij}^{JT}$ are two body matrix elements. For $i,j\equiv r$,
no defects can be detected until much higher in the $pf$ region. On
the contrary, the calculations are quite sensitive to changes in
$W_{ffff}^{JT}$\/ but the only ones that are compulsory affect the
centroids and it is the binding energies that are sensitive to them:

\begin{eqnarray}
  V_{ff}^0{\mbox{(KB1)}}& =& V_{ff}^0{\mbox{(KB)}}-350\,\mbox{
    keV}\nonumber \\ V_{ff}^1{\mbox{(KB1)}}& =&
  V_{ff}^1{\mbox{(KB)}}-110\,\mbox{ keV}\,.
\end{eqnarray}

The interaction we use in the paper, KB3, was defined in
ref.~\cite{pozu2}
as

\begin{eqnarray}
W_{ffff}^{J0}{\mbox{(KB3)}}&=&W_{ffff}^{J0}
{\mbox{(KB1)}}-300\,{\mbox{keV~for~}}J=1,3\nonumber \\
W_{ffff}^{21}{\mbox{(KB3)}}&=&W_{ffff}^{21}
{\mbox{(KB1)}}-200\,{\mbox{keV.}}
\label{eq:kb3}
\end{eqnarray}
while the other matrix elements are modified so as to keep the
centroids (4).

It should be understood that the minimal interaction is KB1 in that
the bad behaviour of the centroids reflects the bad saturation
properties of the realistic potentials: if we do not accept
corrections to the centroids we have no realistic interaction.
Compared with the statements in eqs. (2) and (4), eq. (5) is very
small talk that could just as well be ignored. However, to indulge in
it is of some interest, as will become apparent in section 10.

In what follows, and unless specified otherwise, we use

\begin{itemize}
\item harmonic oscillator wave functions with $b=1.93\,$fm
\item bare electromagnetic factors in $M1$ transitions; effective
  charges of 1.5~e for protons and 0.5~e for neutrons in the electric
  quadrupole transitions and moments.
\item Gamow-Teller (GT) strength defined through
\begin{equation}
    B(GT)=\kappa^2\langle \sigma\tau\rangle^2, \hspace{1cm}
    \langle \sigma\tau\rangle =\frac{\langle \mbox{f}||\sum_k\sigma^k
      t^k_\pm ||\mbox{i}\rangle}{\sqrt{2J_i+1}},
\end{equation}
where the matrix element is reduced with respect to the spin operator
only (Racah convention~\cite{edmonds}) and $\kappa$ is the axial to
vector ratio for GT decays.
\begin{equation}
  \kappa=(g_A/g_V)_{\mbox{\scriptsize
      eff}}=0.77(g_A/g_V)_{\mbox{\scriptsize bare}}=0.963(7).
\end{equation}
\item for the Fermi decays we have
    \begin{equation}
     B(F)=\langle\tau\rangle^2, \hspace{1cm}
   \langle\tau\rangle =\frac{\langle \mbox{f}||\sum_k
      t^k_\pm ||\mbox{i}\rangle}{\sqrt{2J_i+1}}
    \end{equation}

\item half-lives, $t$, are found through

\begin{equation}
(f_A+f^\epsilon)t=\frac{6170\pm4}{(f_V/f_A)B(F)+B(GT)}
\end{equation}

\end{itemize}

We follow ref.~\cite{wilki} in the calculation of the $f_A$ and $f_V$
integrals and ref.~\cite{bamby} for $f^\epsilon$. The experimental
energies are used.

\section{$^{48}$Ca \hspace{1cm} ERRATUM}

\begin{figure}[t]
  \epsffile{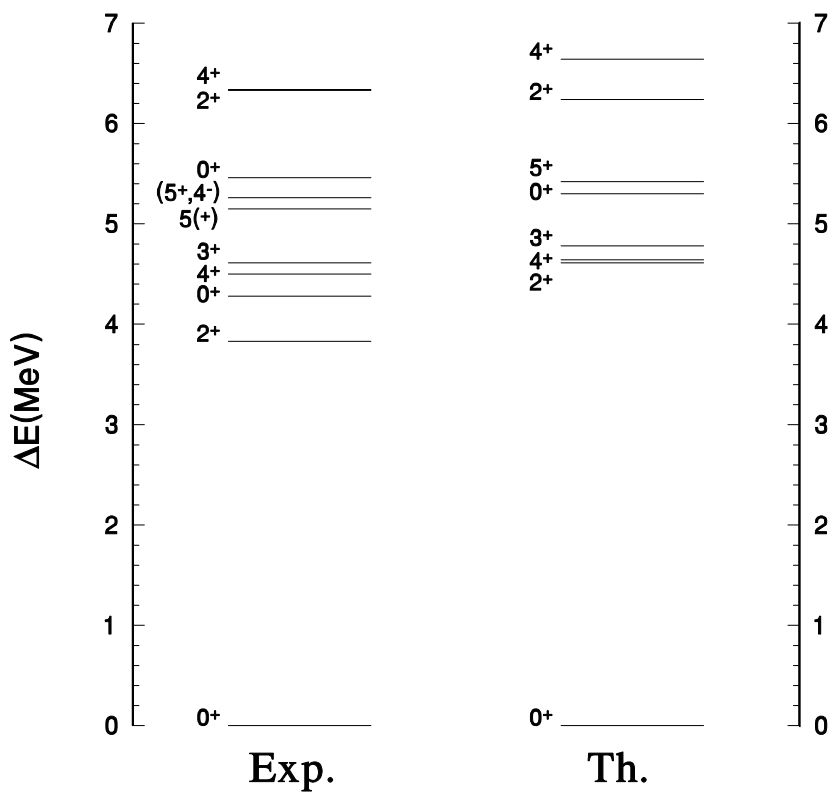}
\caption{\label{fig:eca48} Experimental and theoretical energy levels
  of $^{48}$Ca}
\end{figure}

In fig.~\ref{fig:eca48} we compare calculated and experimental levels.
Except for the lowest $2^+$ state the agreement is good and the first
excited $0^+$ is certainly an intruder.

The calculated $M1$ strength is found in a triplet (not shown) at
excitation energies of 9.82, 10.06 and 10.23$\,$MeV with $B(M1)$ values
of 0.39, 6.17 and 1.81$\,\mu_N^2$, respectively, which nearly exhaust
the sum rule of 8.96$\,\mu_N^2$. The total observed strength between
7.5 and 12.5 MeV is $5.2\pm0.5\ \mu_N^2$. It is dominated by the
majestic peak at 10.2$\,$MeV ($3.9\ \mu_N^2$) and otherwise fragmented
among some twenty states; below 11.7 MeV there are 14 peaks where the
calculation only produce 8. The observed to calculated ratio
$5.2/8.96=(0.76)^2$, is very much the standard value for spin-like
operators.

The $E2$ rates

\begin{eqnarray}
B(E2,4^+\;\rightarrow\;2^+)&=&2.65\,\mbox{e$^2$ fm$^4$}\nonumber \\
B(E2,2^+\;\rightarrow\;0^+)&=&10.2\,\mbox{e$^2$ fm$^4$}
\end{eqnarray}
agree reasonably with the experimental values
\begin{eqnarray}
B(E2,4^+\;\rightarrow\;2^+)_{\mbox{\scriptsize
      exp}}&=&1.89\,\mbox{e$^2$ fm$^4$}\nonumber \\
B(E2,2^+\;\rightarrow
  \;0^+)_{\mbox{\scriptsize exp}}&=&17.2\,\mbox{e$^2$ fm$^4$}
\end{eqnarray}
but definitely suggest that something is missing in a strict
$0\hbar\omega$ calculation.

In ref.~\cite{cpz2} we studied $2\nu$ double $\beta$ decay of $^{48}$Ca
and calculated the strength functions for the associated processes
$^{48}$Ca$(p,n){}^{48}$Sc and $^{48}$Ti$(n,p){}^{48}$Sc, and for the
latter we have to offer the following:

ERRATUM\@. The total $^{48}$Ti$(n,p)$ strength in ref.~\cite{cpz2}
misses a factor 3/2. Hence the $^{48}$Ca
$2\nu$ double $\beta$ decay half-life has to be multiplied by a
factor 2/3 to yield $T_{1/2}=3.7\cdot 10^{19}\,$yr.

\section{$^{48}$Sc The famous seven}

\begin{figure}[t]
  \epsffile{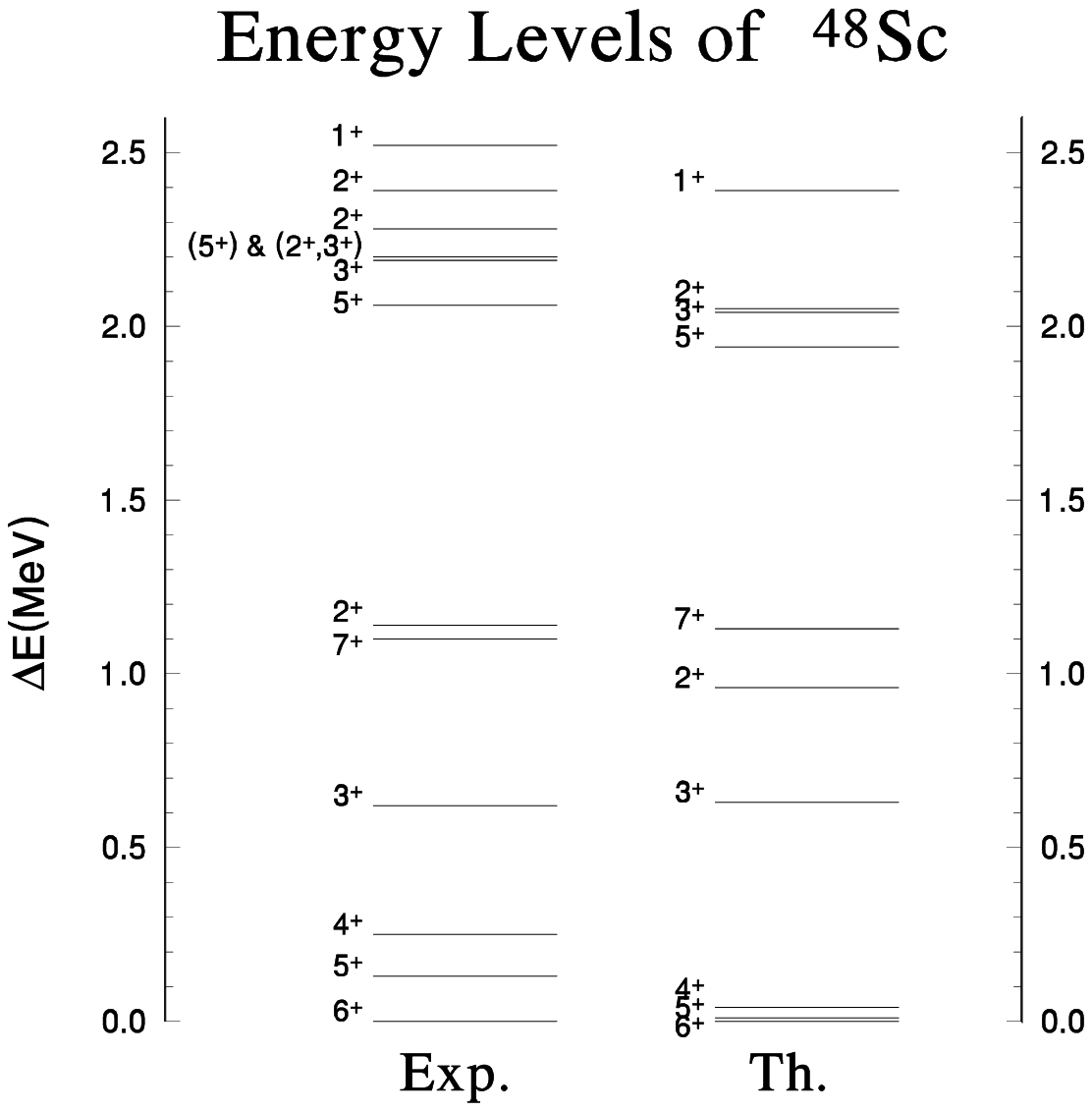}
\caption{\label{fig:esc48} Experimental and theoretical energy levels
  of $^{48}$Sc}
\end{figure}

The $J=1-7$, $T=3$, multiplet of ``$f^8$'' states in $^{48}$Sc can be
related to ``$f^2$'' states in $^{42}$Sc through a Racah coefficient
(the Pandya-Talmi transformation). The operation is successful enough
to have become a textbook example. There are some discrepancies that
are removed by a t=1 truncation, hailed in ref.~\cite{pazu2} as a
triumph of the realistic forces. These levels change very little in
going from t=1 to perturbative~\cite{pozu1,pozu2} and then to the
exact results, but they are quite sensitive to changes in the
$W_{ffff}$ matrix elements. In all the other nuclei the situation is
reversed and figure~\ref{fig:esc48} provides the {\em only\/} example
of an exact calculation that does not bring an improvement over the
approximate ones. Note however that the agreement with data is
definitely good, and extends to three levels at around 2~MeV that do
not belong to the multiplet. Below 2.5~MeV there are a couple of 2$^+$
states with no calculated counterparts, i.e.\ intruders. A more
complete view of the density of intruders comes from
table~\ref{tab:sc48}, where we have listed the 9 calculated 1$^+$
levels below 6~MeV against 13 experimental candidates~\cite{flem}.
Immediately above 6~MeV, diagonalizations yield level spacings of
100~keV. The number of intruders will also grow fast and one to one
identifications become meaningless because the number of levels that
can be observed becomes small fraction of those present.  Note that
even some of the calculated states in table~\ref{tab:sc48} may have
escaped detection (e.g.\ the one at 5.23~MeV).

\begin{table}[h]
  \begin{center}
    \leavevmode
    \begin{tabular}{cccccccccccccc}
      \hline \hline Exp & 2.52 & 2.98 & 3.06 & 3.16 & 3.26 & 3.71 &
      4.00 & 4.18
      & 4.32 & 4.68 & 4.78 & 5.45 & 5.74\\ Th & 2.39 & 2.91 & & & 3.46
      & &
      3.95 & & 4.49 & 4.67 & 5.23 & 5.49 & 5.79\\
\hline\hline
 \end{tabular}
\end{center}
 \caption{1$^+$ states in $^{48}$Sc. Energies in MeV.}
 \label{tab:sc48}
\end{table}

$^{48}$Sc decays to $^{48}$Ti {\it via} the doublet of $6^+$ states at
3.33 and 3.51~MeV. The measured half-life, $\log ft$ values and
branching ratios are

\begin{equation}
\begin{array}{c}
  \mbox{T}_{1/2}=43.7\mbox{ h}\\[0.2cm]
\begin{array}{ll}
  \log ft(6^+_1)= 5.53 & \%\beta=90.7\%\\[0.2cm] \log ft(6^+_2)= 6.01 &
  \%\beta=9.3\%
\end{array}\\
\end{array}
\end{equation}
while the calculated ones read

\begin{equation}
\begin{array}{c}
  \mbox{T}_{1/2}=29.14\mbox{ h}\\[0.2cm]
\begin{array}{ll}
  \log ft(6^+_1)= 5.34 & \%\beta=96\%\\[0.2cm] \log ft(6^+_2)= 6.09 &
  \%\beta=4\%
\end{array}\\
\end{array}
\end{equation}

We have here a first example of the extreme sensitivity of the
half-lives to effects that are bound to produce very minor changes in
other properties that are satisfactorily described, such as those of
the $6^+$ doublet (see next section).

\section{$^{48}$Ti Intrinsic states}

\begin{figure}[h]
  \epsfxsize=14.5cm
  \begin{center}
    \leavevmode
  \epsffile{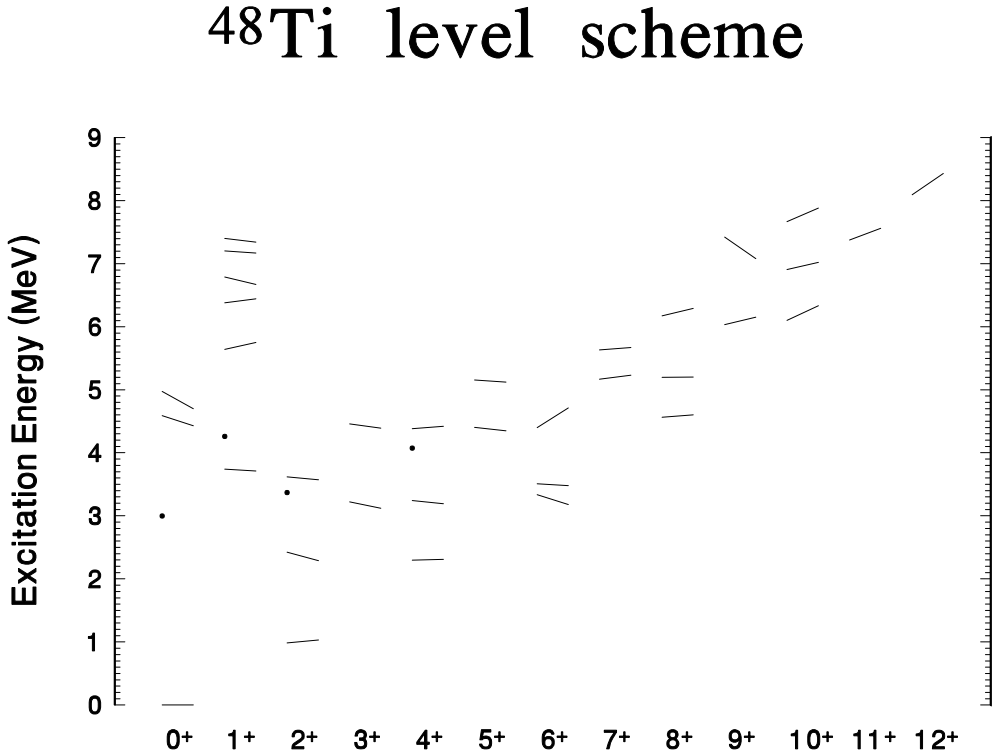}
  \end{center}
\caption{\label{fig:eti48} Experimental and theoretical energy level
  of $^{48}$Ti. The left side of the lines corresponds to the
  experimental value, the right side corresponds to the theoretical
  one.  The isolated points are intruders states not included in our
  valence space}
\end{figure}

Experimentally this is the richest of the A~=~48 nuclei. The lines in
fig.~\ref{fig:eti48} connect theoretical levels (to the right) with
observed ones (to the left). The agreement, which would be perfect if
all lines were horizontal, is nevertheless quite good: the rms
deviation for the 33 excited states is of $110\,$keV. Dots correspond
to the first intruders detected for a given $J$ value.

\clearpage

Tables~\ref{tab:e2ti48} and~\ref{tab:m1ti48} contain the information
on $E2$ and $M1$ transitions, to which we may add the magnetic moment
of the first $2^+$ state, calculated to be $\mu(2^+)=0.43\,\mu_N$,
lowish with respect of the measured $\mu(2^+)=0.86\,(38)\,\mu_N$ and
$\mu(2^+)=1.12\,(22)\,\mu_N$~\cite{radga}.

\begin{table}[h]
  \begin{center}
    \leavevmode
    \begin{tabular}{ccccc}
      \hline \hline
      $J^\pi_n$(i) & $J^\pi_m$(f) & EXP & $(fp)^8$ & $f^8(^{42}$Sc)\\
      \hline
      2$^+_1$ & 0$^+_1$ & 142 $\pm$ 6 & 92.1 & 52.9 \\
      4$^+_1$ & 2$^+_1$ & 111 $\pm$ 20 & 136.6 & 65.8 \\
      2$^+_2$ & 2$^+_1$ & 104 $\pm$ 104 & 36.0 & 45.2 \\
      2$^+_2$ & 0$^+_1$ & 13 $\pm$ 2 & 20.3 & 6.6 \\
      2$^+_3$ & 0$^+_1$ & 22 $\pm$ 6 & 4.3 & 0.2 \\
      6$^+_1$ & 4$^+_1$ & 52 $\pm$ 5 & 54.8 & 39.3 \\
      6$^+_2$ & 4$^+_1$ & 56 $\pm$ 25 & 55.5 & 0.7 \\
      6$^+_3$ & 4$^+_1$ & 75 $\pm$ 30 & 34.1 & 29.8 \\
      8$^+_1$ & 6$^+_2$ & $<$14.5 & 20.2 & 4.6 \\
      8$^+_1$ & 6$^+_1$ & $<$51.8 & 63.4 & 41.4 \\
      8$^+_2$ & 6$^+_2$ & 74 $\pm$ 40 & 46.6 & 32.4 \\
      8$^+_2$ & 6$^+_1$ & 9$^{+9}_{-4}$ & 10.7 & 2.2 \\
      10$^+_1$ & 8$^+_1$ & $<$46.6 & 46.7 & 32.9 \\
      9$^+_2$ & 9$^+_1$ & $<$39.4 & 11.8 &11.7 \\
\hline \hline
    \end{tabular}
  \end{center}
  \caption{$^{48}$Ti $E2$ transitions, $B(E2)$ in e$^2$~fm$^4$.
    $f^8(^{42}$Sc) is a t=0 calculation with $W_{ffff}$ matrix
    elements taken from the spectrum of $^{42}$Sc.}
  \label{tab:e2ti48}
\end{table}

\begin{table}[h]
  \begin{center}
    \leavevmode
    \begin{tabular}{ccccc}
      \hline \hline $J^\pi_n$(i) & $J^\pi_m$(f) & EXP & $(fp)^8$ &
      $f^8(^{42}$Sc) \\ \hline
      0$^+_1$ & 1$^+_1$ & 0.50 $\pm$ 0.08 & 0.54 & 0.73 \\
      0$^+_1$ & 1$^+_2$ & 0.50 $\pm$ 0.08 & 0.41 &  \\
      0$^+_1$ & 1$^+_5$ & 0.80 $\pm$ 0.06 & 0.42 &  \\
      2$^+_2$ & 2$^+_1$ & 0.5 $\pm$ 0.1 & 0.55 & 1.63 \\
      3$^+_1$ & 2$^+_1$ & $<$0.01 & 0.55 & 1.63 \\%  1.08 \\
      3$^+_1$ & 4$^+_1$ & $<$0.06 & 0.39 & 0.82 \\% 0.88 \\
      4$^+_2$ & 4$^+_1$ & 1.0 $\pm$ 0.2 & 1.78 & 3.45 \\
      2$^+_3$ & 2$^+_1$ & 0.05 $\pm$ 0.01 & 0.39 & 0.49 \\
      6$^+_2$ & 6$^+_1$ & 6.7 $\pm$ 3.3 & 2.32 & 8.96 \\
      6$^+_3$ & 6$^+_1$ & 0.9 $\pm$ 0.4 & 0.06 & 0.00 \\
      5$^+_1$ & 6$^+_1$ & $>$0.36 & 0.41 & 1.00 \\
      5$^+_2$ & 6$^+_2$ & $>$0.9 & 0.27 & 0.68 \\
      7$^+_1$ & 6$^+_2$ & 0.09$^{+0.09}_{-0.04}$ & 0.15 & 0.30 \\
      7$^+_1$ & 6$^+_1$ & 0.14$^{+0.14}_{-0.06}$ & 0.33 & 0.00 \\
      8$^+_2$ & 8$^+_1$ & 1.61 $\pm$ 0.70 & 1.59 & 3.38 \\
      7$^+_2$ & 8$^+_1$ & 0.5$^{+0.8}_{-0.2}$ & 0.82 & 0.00 \\
      7$^+_2$ & 6$^+_1$ & 0.07$^{+0.10}_{-0.03}$ & 0.12 & 0.57\\
      9$^+_1$ & 8$^+_2$ & $>$0.95 & 0.82 & 0.00 \\
      9$^+_1$ & 8$^+_1$ & $>$0.36 & 0.57 & 1.55 \\
      8$^+_3$ & 7$^+_1$ & 0.3$^{+1.3}_{-0.1}$ & 0.54 & 1.09 \\
      8$^+_3$ & 8$^+_1$ & 0.2$^{+0.5}_{-0.1}$ & 0.25 & 0.00 \\
      10$^+_2$ & 9$^+_1$ & 0.6$^{+1.0}_{-0.3}$ & 3.60 & 7.08 \\
      11$^+_1$ & 10$^+_2$ & $>$0.36 & 2.58 & 5.37 \\
      11$^+_1$ & 10$^+_1$ & $>$0.36 & 0.81 & 0.00 \\
      12$^+_1$ & 11$^+_1$ & 0.5 $\pm$ 0.2 & 2.32 & 3.46 \\
      \hline \hline
    \end{tabular}
  \end{center}
  \caption{$^{48}$Ti $M1$ transitions, $B(M1)$ in $\mu_n^2$. See also
    caption to table 3.}
  \label{tab:m1ti48}
\end{table}

There are very few discrepancies between the exact calculations
$(fp)^8$, and the data and they are hardly significant. Remarkably
enough this could also be said of the $f^8$ calculations, for the $M1$
transitions and to some extent for the $B(E2)$ with the spectacular
exception in $B(E2,6_{1,2}^+ \rightarrow  4_1^+)$. The exact results
build enough quadrupole coherence with standard effective charges but
the truly important difference with the t=0 calculations shows in
table~\ref{tab:qti48}, where we have collected the intrinsic
quadrupole moment $Q_0$, extracted from the spectroscopic ones through

\begin{equation}
\label{eq:q0qsp}
Q_0=\frac{(J+1)\,(2J+3)}{3K^2-J(J+1)}\,Q_{spec}(J),
\end{equation}
setting $K=0$. For the exact calculations, $Q_0$ is also extracted
through the rotational model prescription

\begin{equation}
\label{eq:q0be2}
B(E2,J\;\rightarrow\;J-2)=\frac{5}{16\pi}\,e^2|\langle JK20|
J-2,K\rangle\, Q_0^2.
\end{equation}

\clearpage

\begin{table}[h]
  \begin{center}
    \leavevmode
    \begin{tabular}{ccccc}
      \hline \hline $J^\pi_n$(i) &  $(fp)^8_s$ & $f^8(^{42}$Sc) &
      $(f_{7/2}p_{3/2})^8$ & $(fp)^8_t$\\ \hline
      2$^+_1$ & 50.4 & $-12.25$ & 39.17 & 68 \\
      4$^+_1$ & 32.18 & $-11.83$ & 18.47 & 69 \\
      6$^+_1$ & $-44.25$ & 13 & 13.99 & \\
      6$^+_2$ & 42 & 4 &  $-25.18$ & 42 \\
      8$^+_1$ & 32.54 & 1.67 & 31.52 & 44\\
      10$^+_1$ & 46.69 & 13.34 & 30.29 & 37 \\
      12$^+_1$ & 35.78 & 28.58 & 31.64 & \\
\multicolumn{5}{c}{$Q_0(2^+)_{\mbox{\scriptsize exp}}=62\pm 3$}\\
      \hline \hline
    \end{tabular}
  \end{center}
  \caption{$^{48}$Ti intrinsic quadrupole moments of the Yrast states
    in e~fm$^{2}$. $(fp)^8_s$ means $Q_0$ extracted from eq.\ (14).
    $(fp)^8_t$ means $Q_0$ extracted from eq.\ (15).}
  \label{tab:qti48}
\end{table}

The wrong sign for the quadrupole moment is a characteristic of the
$t=0$ calculations. The exact results check well with the known
experimental value for the $2^+$ but they do much more: the curious
mixture of signs and sizes coming out of $f^8$, becomes a fairly large
and constant number for $(fp)^8$.

Although we cannot speak of rotational motion ---which demands a truly
constant $Q_0$ and a $J(J+1)$ spectrum--- {\it we are certainly in the
presence of a well defined prolate intrinsic structure}.

This build-up of quadrupole coherence is almost entirely due to mixing
with the $p_{3/2}$ orbit, as seen from the results for the
$(f_{7/2}\,p_{3/2})^8$ space.

\begin{figure}[t]
  \begin{center}
    \leavevmode
  \epsfxsize=13cm
  \epsffile{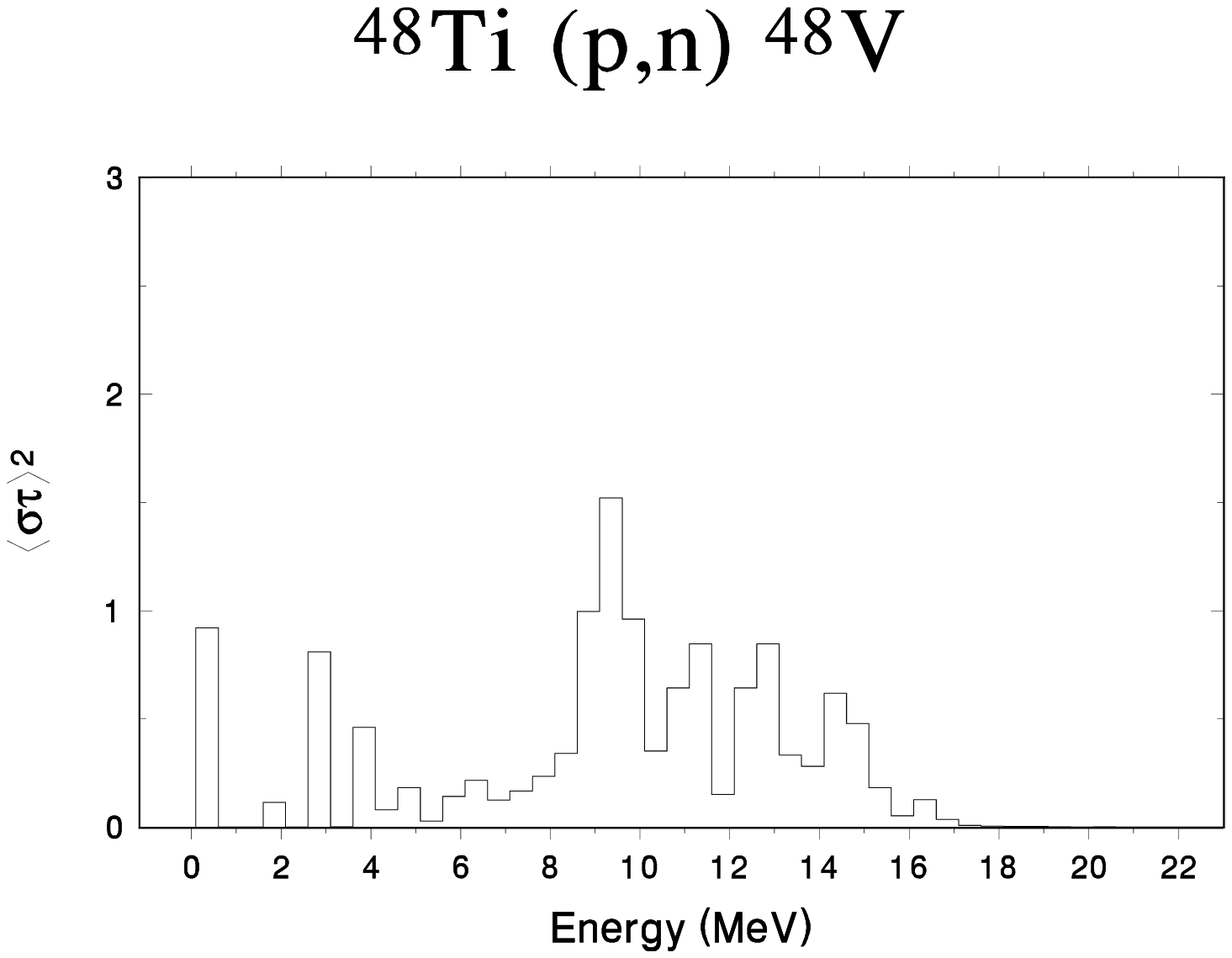}
  \caption{\label{fig:tin1} $^{48}$Ti $\rightarrow$ $^{48}$V
    strength function.\ t~=~1 calculation}

    \leavevmode
    \epsfxsize=13cm
    \epsffile{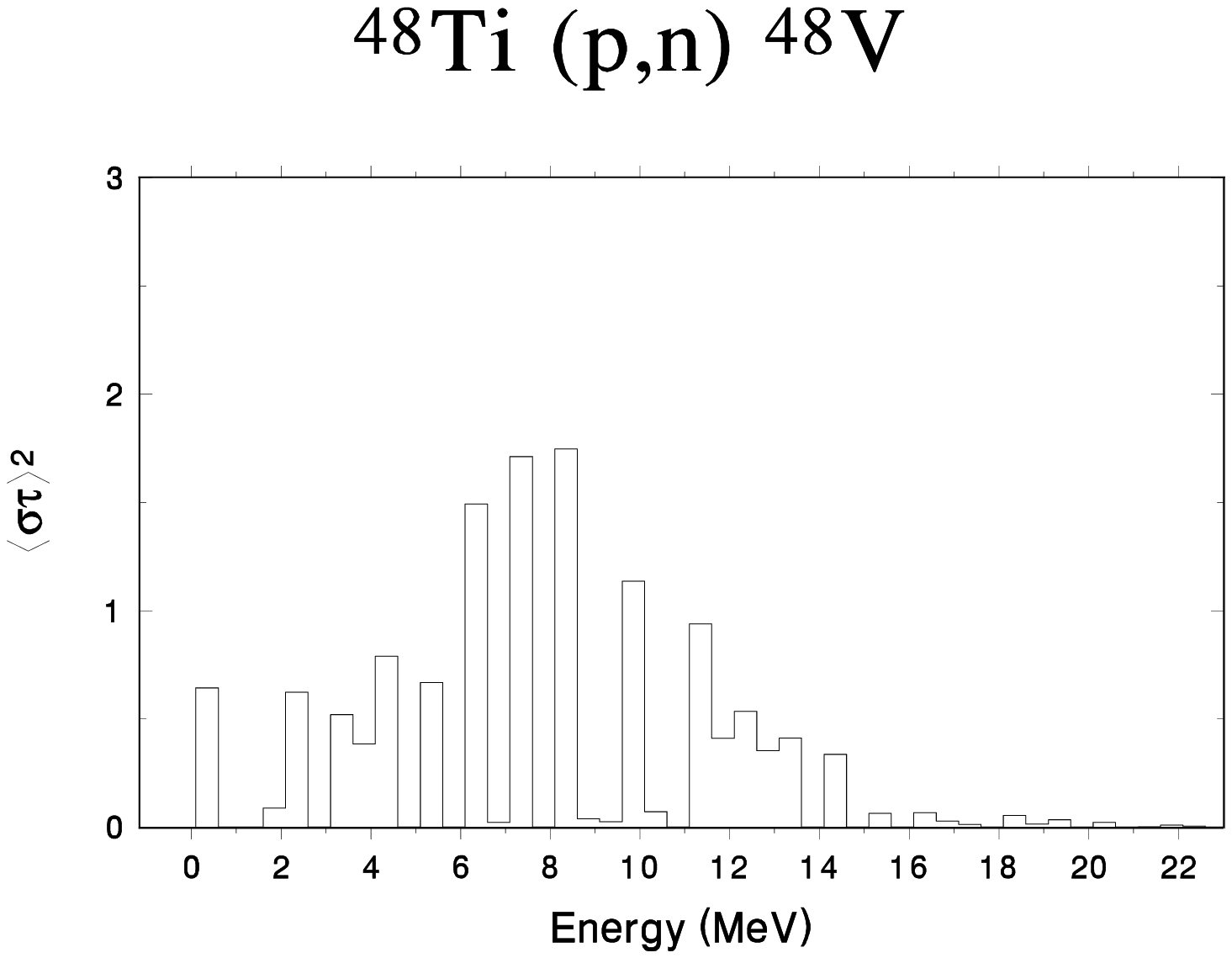}
 \end{center}
\caption{\label{fig:tito} $^{48}$Ti $\rightarrow$ $^{48}$V
  strength function. Full fp-shell calculation}
\end{figure}

The strength function for $^{48}$Ti$(n,p){}^{48}$Sc can be found in
ref.~\cite{cpz2} (remember the erratum though\ldots) and
ref.~\cite{cpz1} contains a study of the $(p,p')$ and $(e,e')$
processes and an analysis of the orbital, spin and $M1$ strengths. The
missing piece of information, $^{48}$Ti$(p,n){}^{48}$V, is found in
figs.~\ref{fig:tin1} and~\ref{fig:tito}, for $t=1$ and exact
calculations. They show similarity in gross structure: some low energy
peaks, a resonance-like middle region and $T=2$ satellite strength
higher up. In details they differ mainly in the position of the
resonance, shifted down by some $2\,$MeV in the exact case. The total
strength for the full space $S^-=13.263\,\kappa^2$, combined with
$S^+=1.263\,\kappa^2$ obtained for $^{48}$Ti$(n,p){}^{48}$Sc,
satisfies the sum rule for the GT operator

\begin{equation}
\label{eq:sumrule}
\sum B(GT^-)-\sum B(GT^+) = S^--S^+=3(N-Z)\,\kappa^2
\end{equation}

The mirror $\beta^+$ decay of $^{48}$Fe to $^{48}$Mn has
$Q_\beta=11.1$~MeV, which covers a large fraction of the strength in
fig.~\ref{fig:tito}. Experimentally, a comparison of the $\beta^+$ and
$(p,n)$ processes seems possible. It would certainly be welcome.

\clearpage

\section{$^{48}$V The interaction}

\begin{figure}[h]
 \epsffile{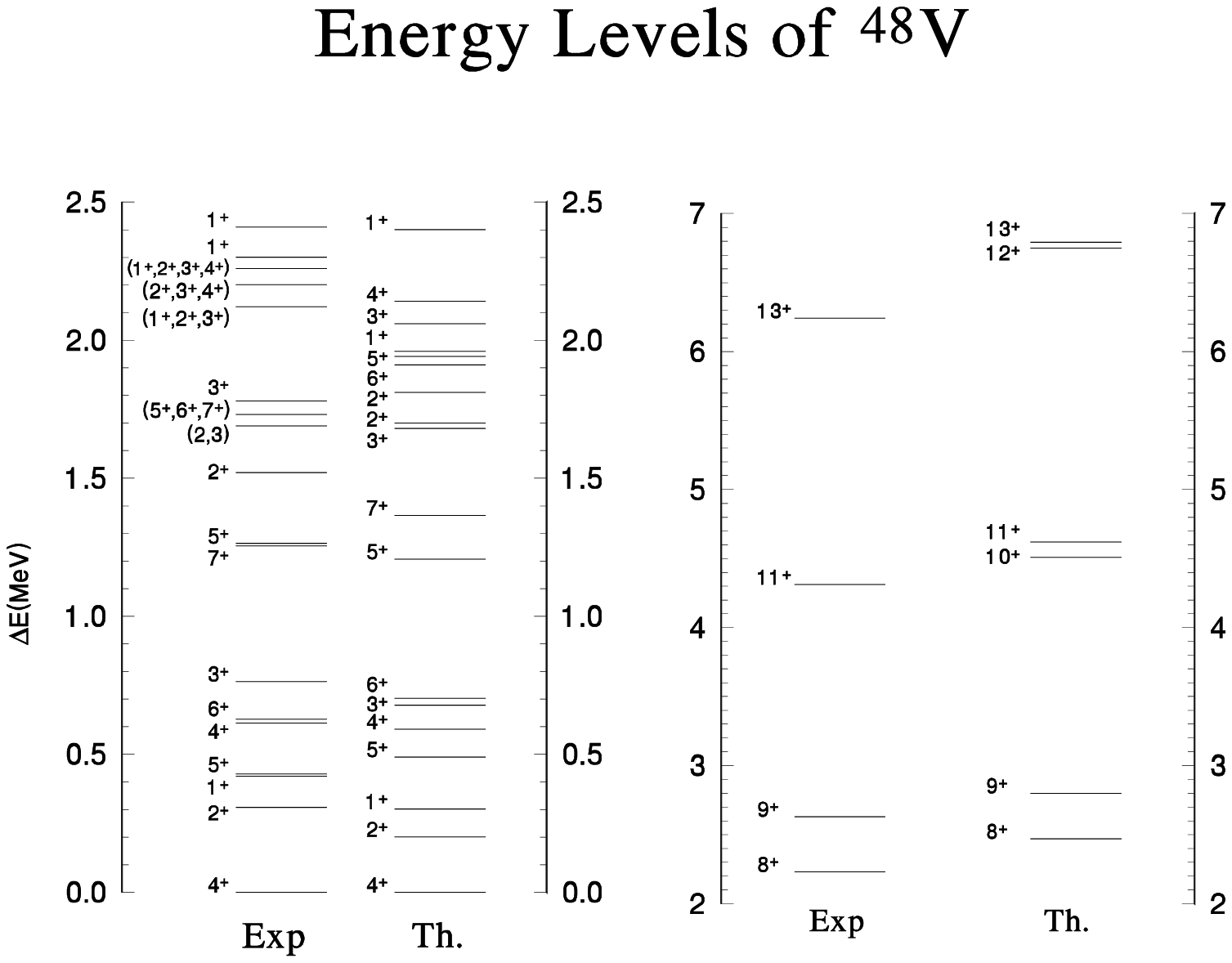}
  \caption{Experimental and theoretical energy levels of $^{48}$V}
  \label{fig:ev48}
\end{figure}

In fig.~\ref{fig:ev48} we have plotted separately all levels with
$J\leq7$ below $2.5\,$MeV and the high spin ones. To the repetitive
claim that the agreement is excellent we may add a mitigating (i.e.\
possibly interesting) remark: it is next to impossible to observe the
calculated $10^+$ and $12^+$ states because of the screening provided
by their close $11^+$ and $13^+$ neighbours.

The $M1$ and $E2$ transitions in table~\ref{tab:emv48} show again
excellent agreement. For the ground state, $\mu(4^+)=1.90\,\mu_N$
against a measured $\mu(4^+)=2.01\,(1)\,\mu_N$, while
$\mu(2^+)=0.51\,\mu_N$ against two measures
$\mu(2^+)=0.28\,(10)\,\mu_N$ and $0.444\,(16)\,\mu_N$~\cite{radga}.

\begin{table}[h]
  \begin{center}
    \leavevmode
    \begin{tabular}{cccc}
      \hline \hline
      $J^\pi_n$(i) & $J^\pi_m$(f) & $B(E2)_{\mbox{\scriptsize exp}}$ &
      $B(E2)_{\mbox{\scriptsize th}}$ \\ \hline
      2$^+_1$ & 4$^+_1$ & 28.59(17) & 48.1 \\
      5$^+_1$ & 4$^+_1$ & 104(42) & 209.0 \\
      4$^+_2$ & 4$^+_1$ & 63(25) &28.9 \\
      4$^+_2$ & 5$^+_1$ & $<$41 & 32.0 \\
      6$^+_1$ & 5$^+_1$ & 186(73) &191.0 \\
      6$^+_1$ & 4$^+_1$ & 46(6) & 52.0 \\
      5$^+_2$ & 4$^+_2$ & $>$176(124) & 41.0 \\
      2$^+_2$ & 2$^+_1$ & $>$1.3(19) & 10.7 \\ \hline
      $J^\pi_n$(i) & $J^\pi_m$(f) & $B(M1)_{\mbox{\scriptsize exp}}$ &
      $B(M1)_{\mbox{\scriptsize th}}$ \\ \hline
      1$^+_1$ & 2$^+_1$ & $>$0.027 & 3.12 \\
      5$^+_1$ & 4$^+_1$ & 0.081(14) & 0.188 \\
      4$^+_2$ & 5$^+_1$ & 0.045(9) & 0.032 \\
      4$^+_2$ & 4$^+_1$ & 0.0084(9) & 0.0079 \\
      6$^+_1$ & 5$^+_1$ & 0.027(5) &0.027 \\
      \hline \hline
    \end{tabular}
  \end{center}
  \caption{$^{48}$V electromagnetic transitions, $B(E2)$ in
    e$^2$~fm$^4$ and $B(M1)$ in $\mu_N^2$}
  \label{tab:emv48}
\end{table}

\begin{table}[h]
  \begin{center}
    \leavevmode
    \begin{tabular}{ccc}
      \hline \hline
      $J_f$ & $B(GT)_{\mbox{\scriptsize th}}$ &
        $B(GT)_{\mbox{\scriptsize exp}}$ \\ \hline
      4$^+_1$ & 7.9 10$^{-3}$ & 4.1 10$^{-3}$ \\
      3$^+_1$ & 0.5 10$^{-3}$ & 1.7 10$^{-3}$ \\
      4$^+_2$ & 2.2 10$^{-3}$ & 4.1 10$^{-3}$ \\
      \hline \hline
    \end{tabular}
  \end{center}
  \caption{$^{48}$V (4$^+$) $\rightarrow$ $^{48}$Ti ($J_f$) beta decay}
  \label{tab:b+v48}
\end{table}

\begin{figure}[t]
  \epsffile{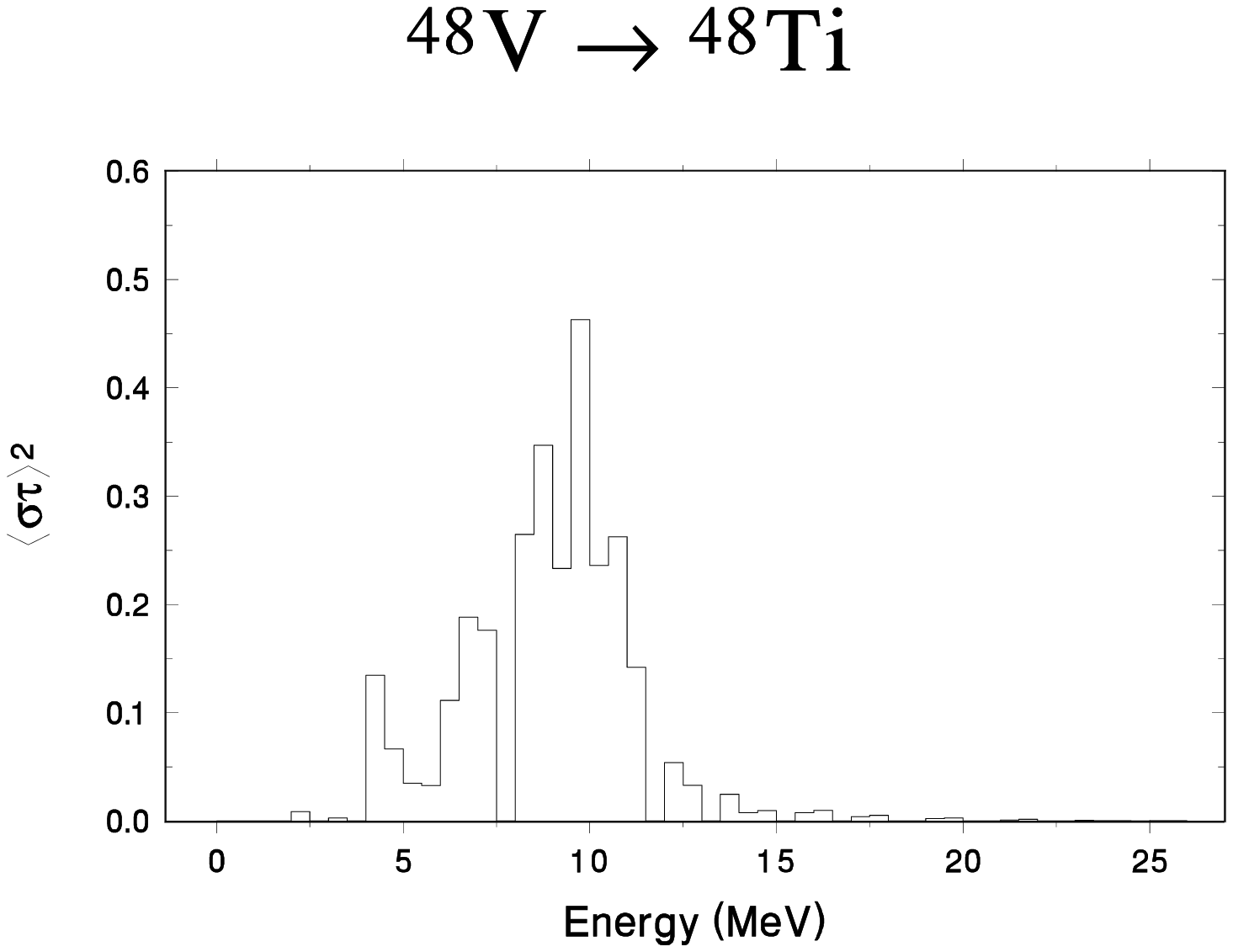}
  \caption{Strength function for the $^{48}$V beta decay. Full $pf$
    shell calculation}
  \label{fig:v48gt}
\end{figure}

The Gamow-Teller $\beta^+$ decay to $^{48}$Ti can reach $J=3,\,4$ and
5 daughter levels and three of them are fed as shown in
table~\ref{tab:b+v48}.  The resulting half life $T_{1/2}=8.85\,$d is
almost half the observed one, $T_{1/2}=15.97\,$d. The fraction of the
total strength responsible for the decay is very small (0.4\%).
Therefore, the discrepancy in the half-lives could be cured by reducing
by half the height of the lowest two bumps in figure~\ref{fig:v48gt},
which is hardly a change in the overall picture.

\clearpage

Odd-odd nuclei provide a good test for the interactions and $^{48}$V
offers a good example of a general trend: perturbative
calculations~\cite{pozu2} are quite good but the exact results come
definitely closer to the data. This systematic improvement clearly
indicates that the interaction is excellent. The one exception to the
trend comes in $^{48}$Sc, and is related to the choice of the KB3
(rather than KB1) interaction. As far as A~=~48 nuclei are concerned
this is an almost irrelevant problem, but there is a conceptual point
that will be made clear in section 10.

\clearpage

\section{$^{48}$Cr Rotational motion}

\begin{figure}[h]
  \epsfxsize=10cm
  \begin{center}
    \leavevmode
       \epsffile{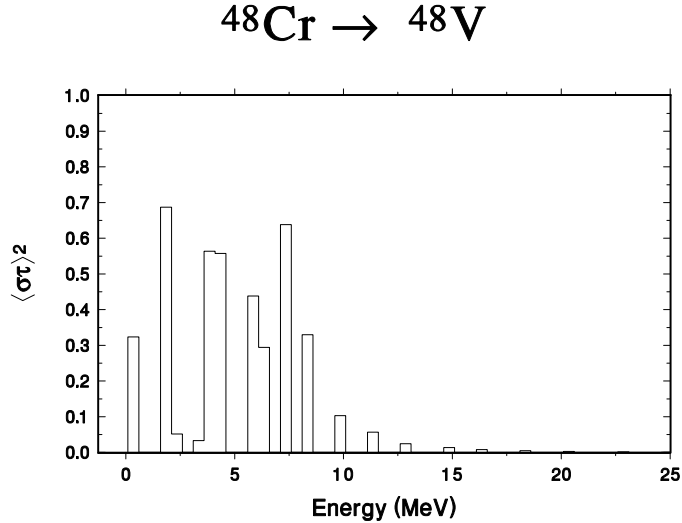}
  \end{center}
 \caption{$^{48}$Cr $\rightarrow$ $^{48}$V strength function.
    Full $pf$ shell calculation}
\label{fig:crva}
\end{figure}

Some levels in $^{48}$Cr are populated in the decay of $^{48}$Mn and
will be discussed in the next section. Beyond that, little is known of
the spectrum, except the Yrast line and before we come to it, we go
through fig.~\ref{fig:crva} for the strength function. Only the first
$1^+$ state is seen through the $Q_\beta$ window. The calculated
half-life is $T_{1/2}=21.8\,$h and the observed one
$T_{1/2}=21.56\,(3)\,$h.

\begin{figure}[t]
  \epsffile{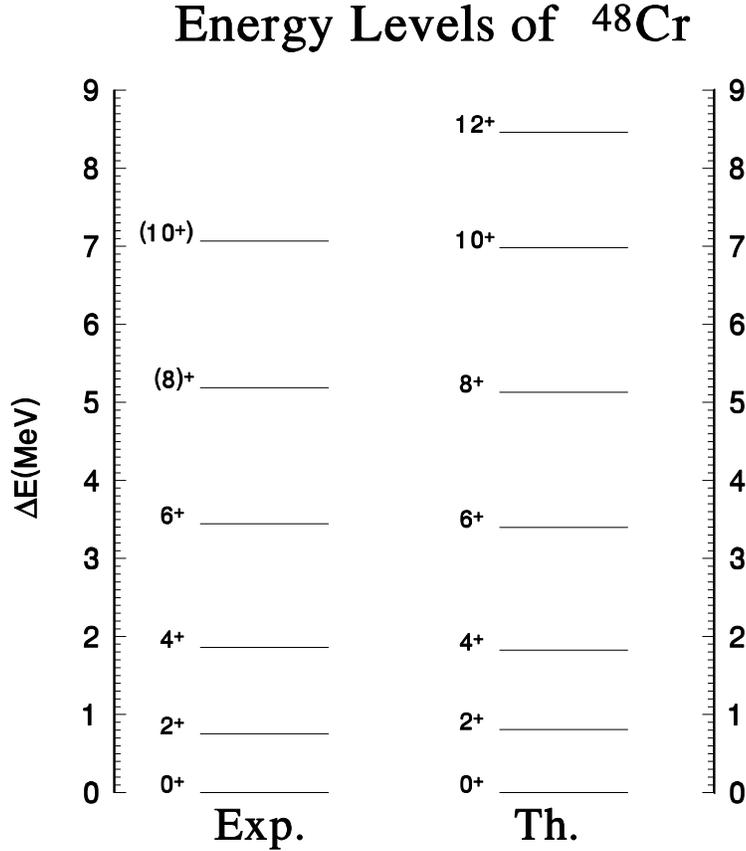}
  \caption{Predicted and experimental Yrast band of $^{48}$Cr}
  \label{fig:ecr48}
\end{figure}

Table~\ref{tab:emcr48} and fig.~\ref{fig:ecr48} collect the
information about the Yrast levels. The only complaint we may have
with the data is the $B(E2,8^+\rightarrow 6^+)$ value but the
calculations are telling us much more than they agree with
observations.

\begin{table}[h]
  \begin{center}
    \leavevmode
    \begin{tabular}{cccccc}
      \hline \hline
     $J$& \multicolumn{2}{c}{$B(E2,J \rightarrow J - 2)$}  &
      $Q_{\mbox{\scriptsize spec}}$ &
      \multicolumn{2}{c}{$Q_{0}$}\\
     &\multicolumn{1}{c}{exp.} &\multicolumn{1}{c}{th.} &
     &\multicolumn{1}{c}{from
     $Q_{\mbox{\scriptsize spec}}$} &\multicolumn{1}{c}{from $B(E2)$}
      \\ \hline
      2 & 321 $\pm$ 41 & 228 & $-29.5$ & 103 & 107 \\
      4& 259 $\pm$ 83 & 312 & $-39.2$ & 108 & 105 \\
      6 &$>$161 & 311 & $-39.7$ & 99 & 100 \\
      8 & 67 $\pm$ 23 & 285 & $-38.9$ & 93 & 93 \\
      10 & $>$35 & 201 & $-22.5$ & 52 & 77 \\
      12 & & 146 & $-5.3$ & 12 & \\ \hline \hline
    \end{tabular}
  \end{center}
  \caption{Electromagnetic properties of the Yrast band of $^{48}$Cr,
    $B(E2)$ in e$^2$~fm$^4$, $Q$ in e~fm$^2$}
  \label{tab:emcr48}
\end{table}

\clearpage

According to the criteria of constancy for $Q_0$ and $J(J+1)$ spacings
defined in section~5 we cannot speak of a {\it good} rotor yet, but we
are coming close to it. The $J(J+1)$ behaviour is at best incipient
but the constancy of $Q_0$ is quite convincing.

Two questions come naturally: what is the mechanism?, can we get
better rotors?

Table~\ref{tab:qti48} suggests that it is the mixing of $f_{7/2}$ and
$p_{3/2}$ orbits that is at the origin of a well defined prolate
intrinsic state. If we follow this hint for any $\Delta j=2$
sequences, it is easy to produce good rotors. The evidence is
presented in table~\ref{tab:rotor}. It contains more information than
necessary to make the point and should be viewed as introductory
material for a future communication. Here we limit ourselves to a few
comments.

We use the bare KLS force~\cite{kls} (all realistic interactions are
very similar~\cite{pazu1,pazu2,acz,duzu} and for this one we have a
code~\cite{lee}). The oscillator frequency is $\hbar\omega=9$, the
spacings between levels are of $1\,$MeV (higher $j$ orbits coming
below) and $Q_0(J)$ is defined in eq. (\ref{eq:q0qsp}).

Under these conditions we obtain truly good rotors. Their stability
has been tested under three types of variation.

\begin{itemize}
\item[i)] To affect all $T=1$ matrix elements by a dilation factor
  $\lambda\approx 1.5$ mocks surprisingly well core polarization
  effects in a major shell~\cite{acz,duzu}. Rotational behaviour
  persists and may be even emphasized.

\item[ii)] For splittings between levels of $2\,$MeV the rotational
  features are eroded but $E_4/E_2>3$ in all cases.

\item[iii)] Pairing is the most efficient enemy of good rotors. To go
  from KLS (bare) to KB (renormalized), $\lambda=1.5)$ (from i)) is
  sufficient, except for $W_{ffff}^{01}$ which should be doubled. Then
  $E_4/E_2\approx 2.5$, not far from what is seen in
  fig.~\ref{fig:ecr48}.
\end{itemize}

\begin{table}[h]
  \begin{center}
    \leavevmode
\renewcommand{\arraystretch}{1.2}
    \begin{tabular}{|ccccccccccc|}
    \hline \hline
      & \multicolumn{2}{|c|}{$(f_{7}p_{3})^8\ T=0$}  &
      \multicolumn{2}{c|}{$(f_{7}p_{3})^4(g_9d_5s_1)^4$}  &
      \multicolumn{2}{c|}{$(g_9d_5s_1)^8\ T=0$}  &
      \multicolumn{2}{c|}{$(g_9d_5s_1)^4(h_{11}f_7p_3)^4$}  &
      \multicolumn{2}{c|}{rigid rotor} \\ \hline
$J$ & $E$ & $Q_0$ & $E$ & $Q_0$ & $E$ & $Q_0$ &
$E$ & $Q_0$ & &
\\ \hline
2 & 0.112 & 133 & 0.043 & 140 & 0.031 & 189 & 0.011 & 193 &  &  \\
4 & 0.365 & 133 & 0.148 & 140 & 0.101 & 189 & 0.034 & 192 & &
\\
6 & 0.756 & 132 & 0.317 & 139 & 0.217 & 189 & 0.075 & 191 &  &  \\
8 & 1.295 & 127 & 0.561 & 137 & 0.385 & 188 & 0.142 & 189 &  &  \\
\hline \hline
& $\frac{E(J)}{E(2^+)}$ & $\frac{Q_0(J)}{Q_0(2^+)}$ &
$\frac{E(J)}{E(2^+)}$ & $\frac{Q_0(J)}{Q_0(2^+)}$ &
$\frac{E(J)}{E(2^+)}$
& $\frac{Q_0(J)}{Q_0(2^+)}$ & $\frac{E(J)}{E(2^+)}$ &
$\frac{Q_0(J)}{Q_0(2^+)}$ & $\frac{E(J)}{E(2^+)}$ &
$\frac{Q_0(J)}{Q_0(2^+)}$ \\[.2cm] \hline
4 & 3.26 & 1 & 3.32 & 1 & 3.44 & 1 & 3.29 & 0.995 & 3.33 & 1 \\
6 & 6.75 & 0.992 & 7.13 & 1 & 7.37 & 0.993 & 6.74 & 0.990 & 7 & 1 \\
8 & 11.56 & 0.955 & 12.66 & 0.995 & 13.04 & 0.979 & 12.72 & 0.979 & 12
& 1 \\
%10 & 17.49 & 0.707 & 20.56 & 0.984 & 20.47 & 0.921 & 21.81 & 0.964 &
%18.33 & 1 \\
%12 & 22.79 & 0.211 & 30.57 & 0.947 & 29.59 & 0.721 & 34.66 & 0.933 &
%26 & 1 \\
\hline \hline
    \end{tabular}
  \end{center}
  \caption{Excitation energies $E$ (in MeV) and intrinsic quadrupole
    moments $Q_0$ (in fm$^2$) for the Yrast states of several $\Delta
    j =2$ configurations. The notation for the orbits is $l_{2j}$.}
  \label{tab:rotor}
\end{table}

The relevance of these results to real rotors comes directly from
Nilsson diagrams which tell us that the onset of deformation at
$N\approx 60$, 90 and 136 corresponds to the sudden promotion of two
pairs of neutrons $(\nu)$ and two pairs of protons $(\pi)$ to $K=1/2$,
3/2 orbits originating in the spherical ones $\nu h_{11/2}/\pi
g_{9/2}$ $\nu i_{13/2}/\pi h_{11/2}$ and $\nu j_{15/2}/\pi i_{13/2}$
respectively. Naturally we identify the intrinsic state of the eight
particles with the one coming out of the diagonalizations, which are
possible right now for the first of the three regions, i.e.\ for the
space $(h_{11}f_7p_3)^4_\nu\ (g_9d_5s_1)^4_\pi$ (m-scheme dimension
2$\cdot 10^6$). In table~\ref{tab:rotor} we also show the results for
$(g_9d_5s_1)^4_\nu\ (f_7p_3)^4_\pi$, which we expect to become active
for Z~$\geq$~20, N~$\approx$~40. The $(g_9d_5s_1)^8$ group is very
likely responsible for rotational behaviour in the A~=~80 region and
may be even a candidate for ``superbands'' at lower masses.

\clearpage

\section{$^{48}$Mn Truncations and GT strength}

Spectroscopically, $^{48}$Mn is identical to $^{48}$V (to within
Coulomb effects). Its decay to $^{48}$Cr covers a non negligible
fraction of the strength function~\cite{roeck1,roeck2}. Since this
process, and similar ones in the region, have been analyzed so far
with $t=1$ calculations, we are going to compare them with $t=3$ and
exact ones.  A digression may be of use.

In a decay, the parent is basically an $f^n$ state. Some daughters are
also of $f^n$ type but most are $f^{n-1}r$. In a $t=1$ calculation,
both configurations are present but $f^n$ is allowed to mix with
$f^{n-1}r$ through $W_{fffr}$ matrix elements while $f^{n-1}r$ is not
allowed to go to $f^{n-2}r^2$. At the $t=2$ level, pairing (i.e.\
$W_{ffrr}$ matrix elements) comes in and pushes $f^n$ down through
mixing with $f^{n-2}r^2$, while $f^{n-1}r$ cannot benefit from a similar
push from $f^{n-3}r^3$. It is only at the $t=3$ level that both $f^n$
and $f^{n-1}r$ states can be treated on equal footing.  Now that $t=3$
calculations are possible up to A~=~58, it is useful to check their
validity.

The results for the half life of $^{48}$Mn are collected in
table~\ref{tab:mncr}. We see that the $t=3$ truncation is quite
acceptable. For the total half-life the $t=1$ number is not too bad if
we remember that full calculations are not always as accurate as they
are here. However this relatively good performance owes much to a
strong Fermi branch: the partial GT value for $t=1$ is simply bad. The
last line of table~\ref{tab:mncr} gives an idea of the consequences of
using calculated instead of observed energies for the transitions.

\begin{table}[h]
  \begin{center}
    \leavevmode
    \begin{tabular}{lcccc}
      \hline \hline
      half-life & $t=1$ & $t=3$ & full & exp \\ \hline
      Total & 99 ms & 133 ms & 142 ms & 158(22) ms \\
      Partial Fermi & 244 ms & 244 ms & 244 ms & 275(20) ms \\
      Partial GT & 107 ms & 292 ms & 340 ms & 372(20) ms \\
      Total (theoretical energies) & 41 ms & 80 ms & 116 ms & \\
      \hline \hline
    \end{tabular}
  \end{center}
  \caption{$^{48}$Mn half-lives with different truncations compared
    with the
    experimental results}
  \label{tab:mncr}
\end{table}

\begin{figure}[t]
  \begin{center}
    \leavevmode
  \epsfxsize=8.5cm
    \epsffile{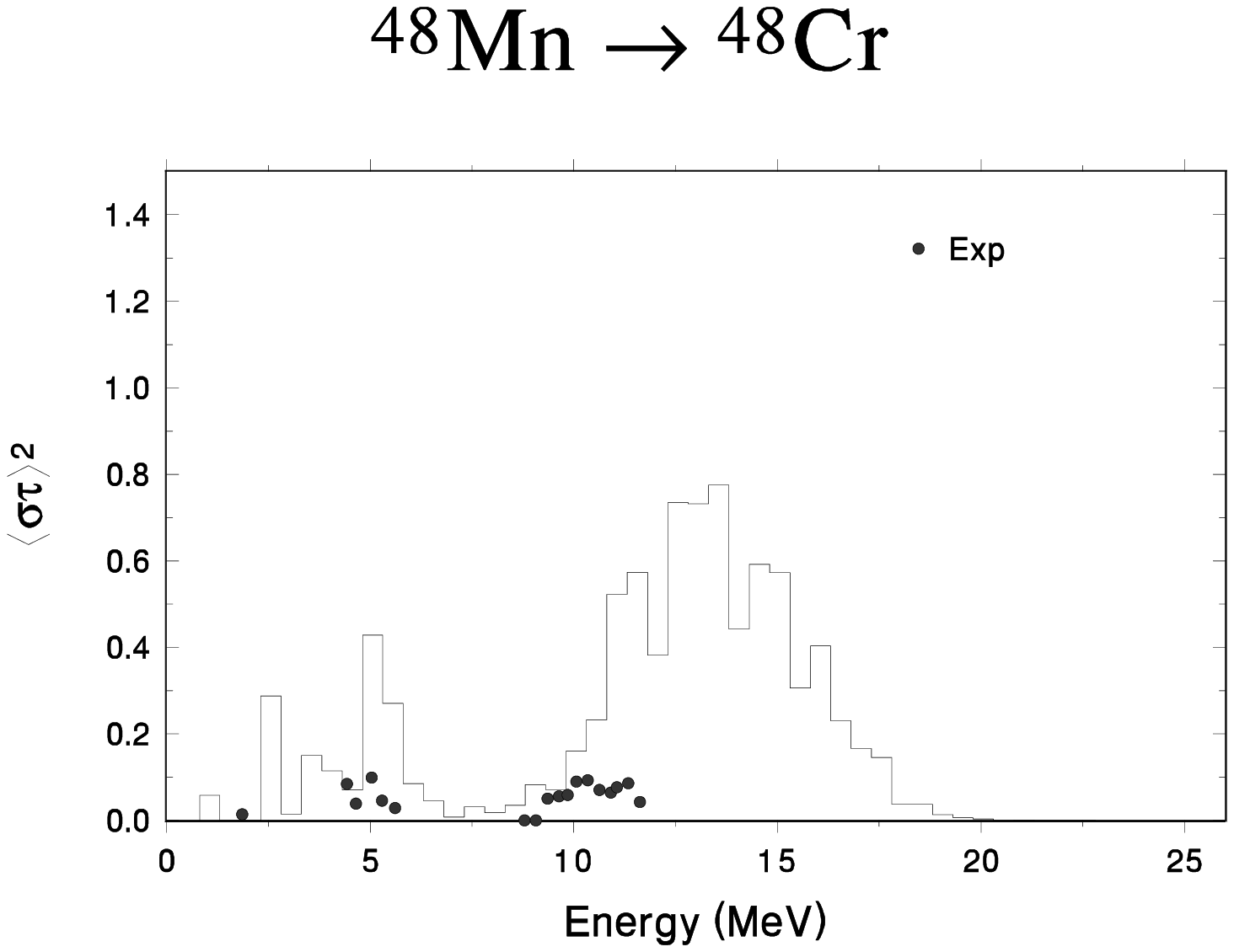}
  \caption{\label{fig:mncr1} $^{48}$Mn $\rightarrow$ $^{48}$Cr
    strength function.\ $t=1$
    calculation}

    \leavevmode
  \epsfxsize=8.5cm
  \epsffile{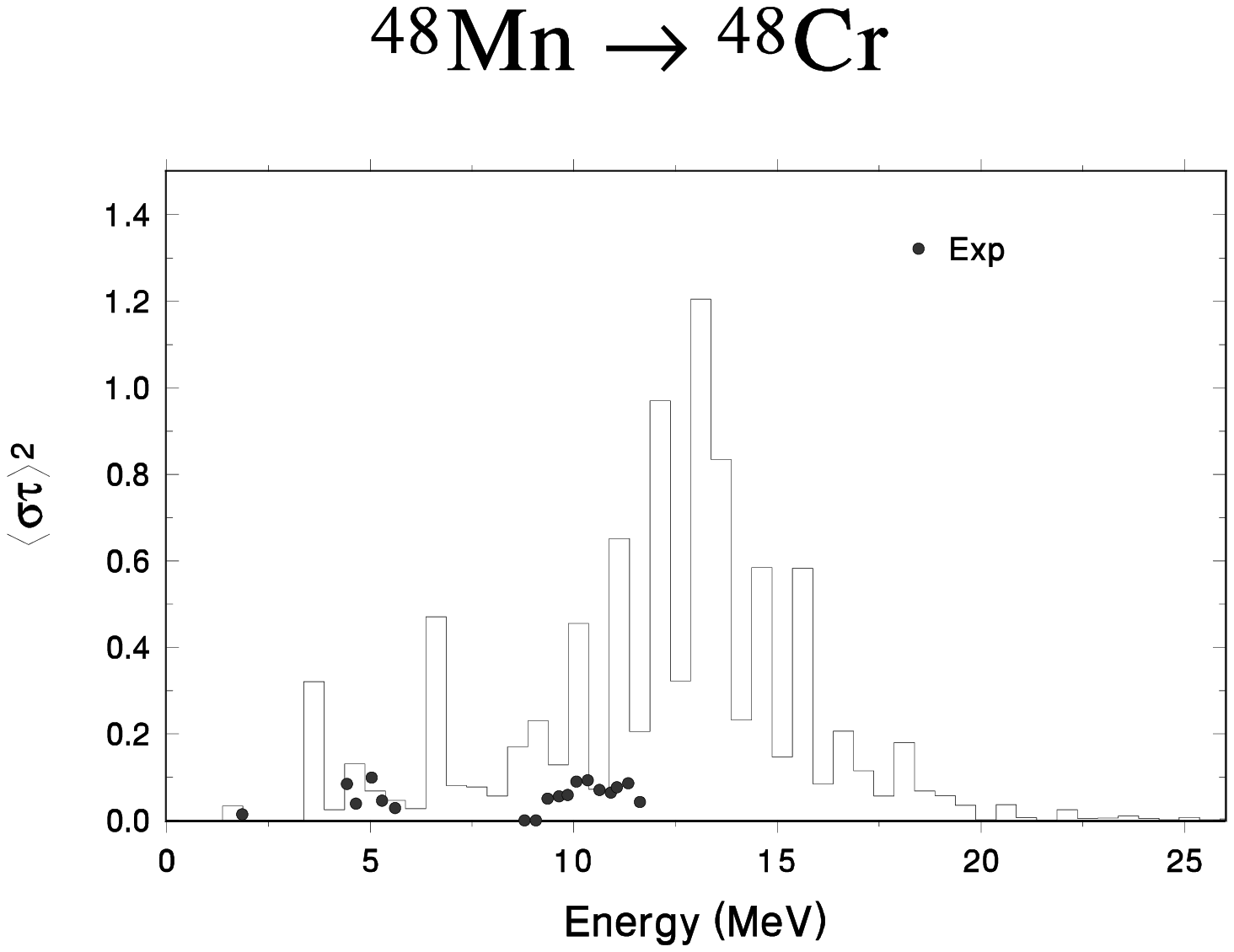}
  \caption{\label{fig:mncr3} $^{48}$Mn $\rightarrow$ $^{48}$Cr
    strength function.\ $t=3$
    calculation}

    \leavevmode
  \epsfxsize=8.5cm
  \epsffile{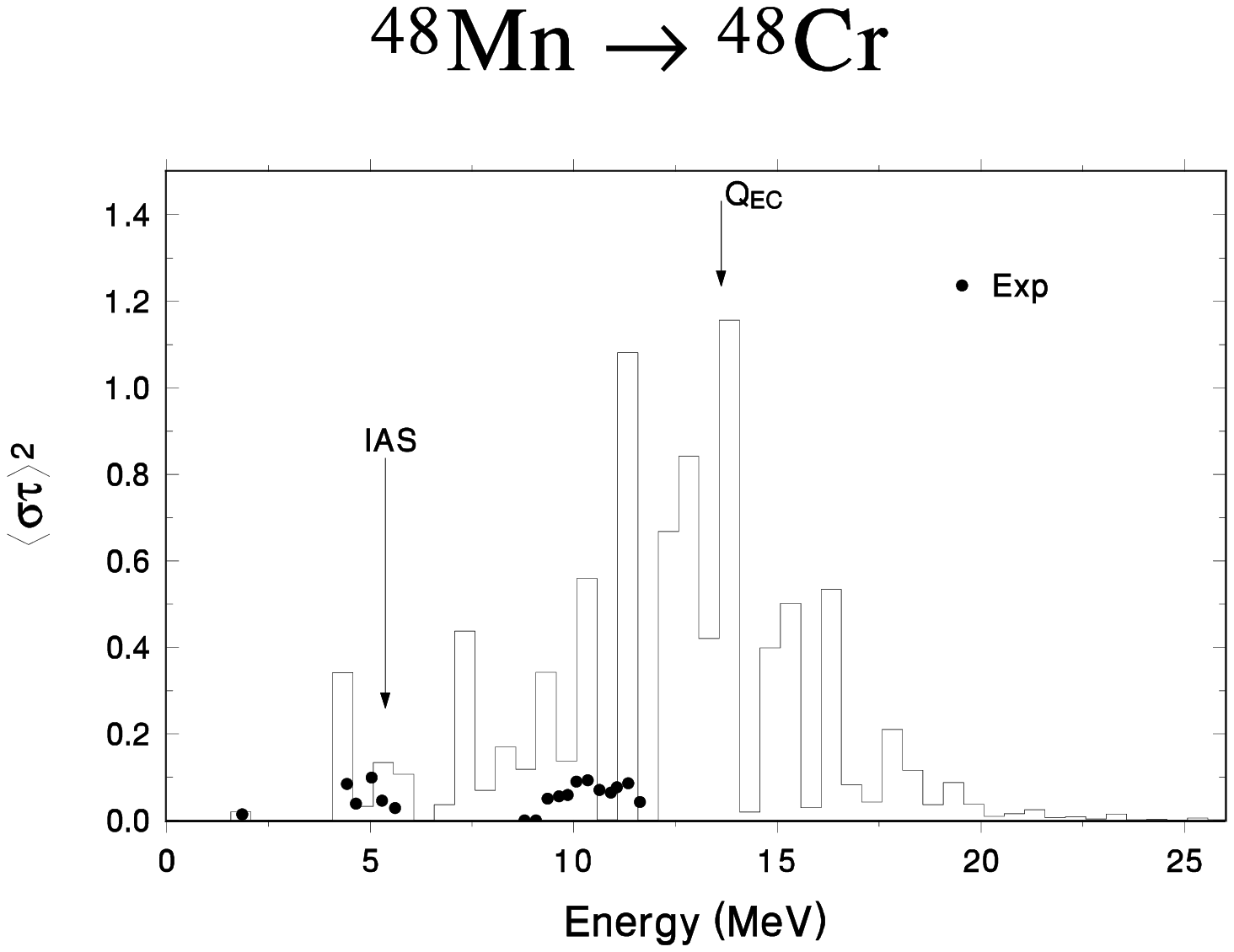}
  \caption{\label{fig:mncrt} $^{48}$Mn $\rightarrow$ $^{48}$Cr
    strength function. Full
    calculation}
  \end{center}
\end{figure}

Next we move to the strength functions in
figures~\ref{fig:mncr1},~\ref{fig:mncr3} and~\ref{fig:mncrt}, where
the data show as dots.  Globally the three calculations agree in that
there is a resonance at $\sim 13\,$MeV and structure at $\sim5\,$MeV.
Beyond that, $t=3$ provides a fairly accurate view of the exact shape
while $t=1$ does not. To give an idea of the energetics: the $4^+\,T=1$
analog (IAS) of the $^{48}$Mn ground state is at $5.79\,$MeV in
$^{48}$Cr, $t=1$ puts it at $3.64\,$MeV, $t=3$ at $4.77\,$MeV and the
exact value is $5.38\,$MeV (not very good by our standards, see later
in section~11).

\begin{figure}[t]
  \epsffile{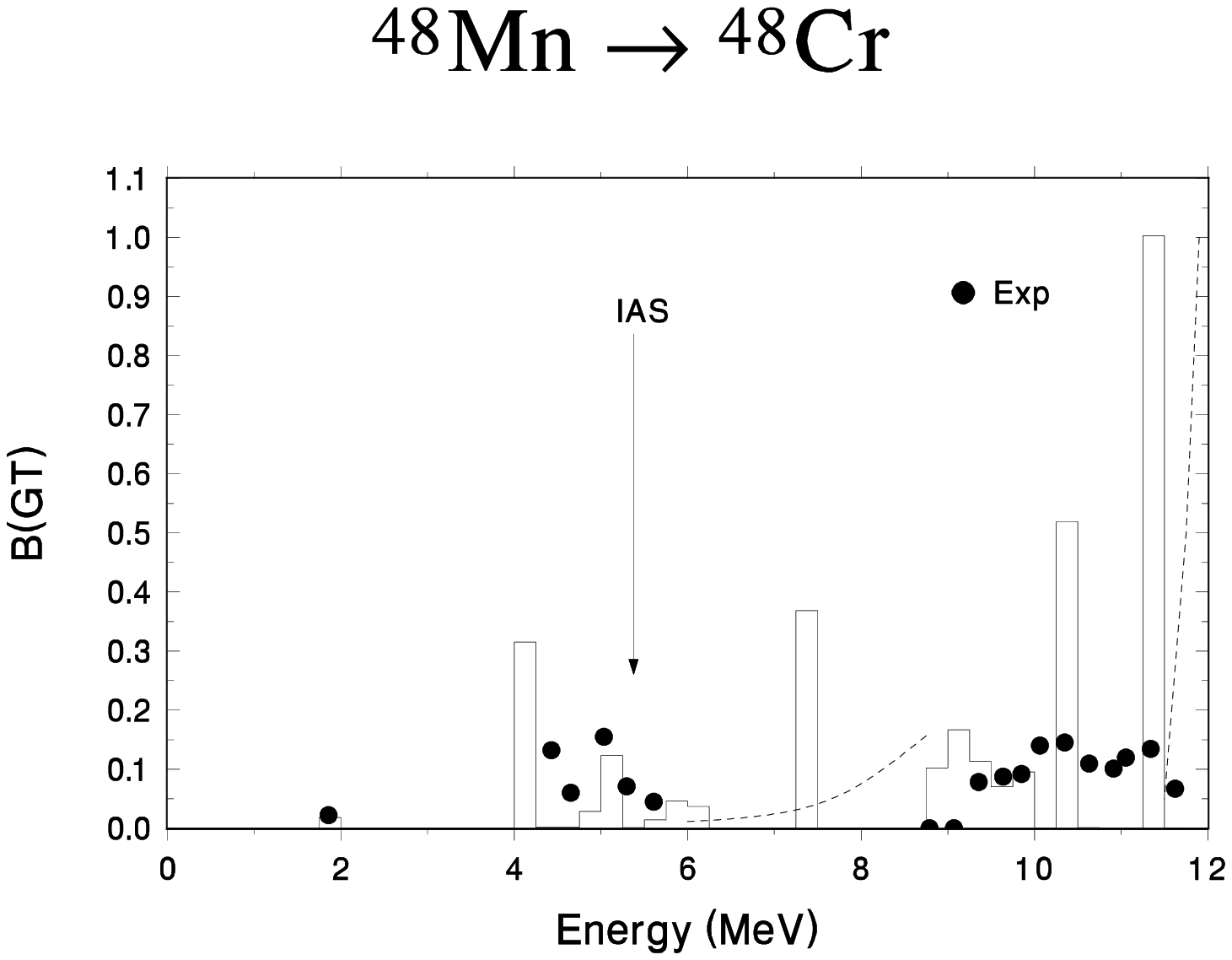}
  \caption{Blow up of the low energy part of figure 12. Here
    we plot $B(GT)$. Theoretical $B(GT)$ values include standard
    $(0.77)^2$ quenching. Dotted lines show the observational
    thresholds [39]. Bins of 250~keV for the calculations (vs.\
    500~keV in figs.~10-11).}
  \label{fig:mncutoff}
\end{figure}

Now we come to the data and concentrate on the exact calculation.
Figure~\ref{fig:mncutoff} contain very much the same information as
figure~\ref{fig:mncrt} but instead of $\langle\sigma\tau\rangle^2$ we
represent $B(GT)$, affected by the $(0.77)^2$ quenching factor for the
calculated numbers. Furthermore, between 6 and 8.5~MeV we have
eliminated among these the peaks that fall below the observation
threshold (shown as a dashed curve)~\cite{roeck2}.

\clearpage

At low energy, the peak at 2~MeV and the cluster of states centered at
5~MeV are very well positioned in the calculations that yield a
strength of 0.504 in the interval [0,5.75]~MeV against an observed
0.485.  Between 0.75 and 8.75~MeV nothing is seen experimentally, and
the only calculated strength above the sensitivity limit is located in
two peaks in the [7.25,7.50]~MeV bin. At these energies the calculated
levels are not yet eigenstates of the system but doorways whose
strength (0.405) will be fragmented (see next section).

Above 8.75~MeV observation resumes through delayed protons yielding
$B(GT)=1.073$ in the [8.75,11.75]~MeV interval versus the calculated
2.07 mostly found in two bins at [10.25,10.50] and [11.25,11.50]~MeV.
At these energies the density of levels is high and fragmentation will
become important. Furthermore, half of the calculated strength is in
the last bin, dangerously close to the abrupt rise in the sensitivity
threshold. Therefore, before we conclude that experiments demand {\em
  anomalous\/} quenching (i.e.: beyond the standard value) we shall
analyse more closely what is being calculated and what is being
measured.

\section{Quenching, shifting and diluting GT strength}

{}From all we have said about GT transitions, a broad trend emerges:
low lying levels are {\em very well\/} positioned and within minor
discrepancies have the observed GT strength once standard quenching is
applied. The examples of good energetics are particularly significant
for the group around 5~MeV in $^{48}$Cr and the 1$^+$ levels in
$^{48}$Ti (fig.~\ref{fig:eti48}) and $^{48}$Sc (table~\ref{tab:sc48}),
that have experimental counterparts within 100~keV more often than
not. The discrepancies are related to the shortish half-life of
$^{48}$Sc and $^{48}$V, a slight lack of spin strength in the lowest
states of $^{48}$Ti (ref.~\cite{cpz1}) and ---perhaps--- with the tail
of the resonant structure in $^{48}$Cr discussed in the preceding
section.

\begin{figure}[h]
  \epsffile{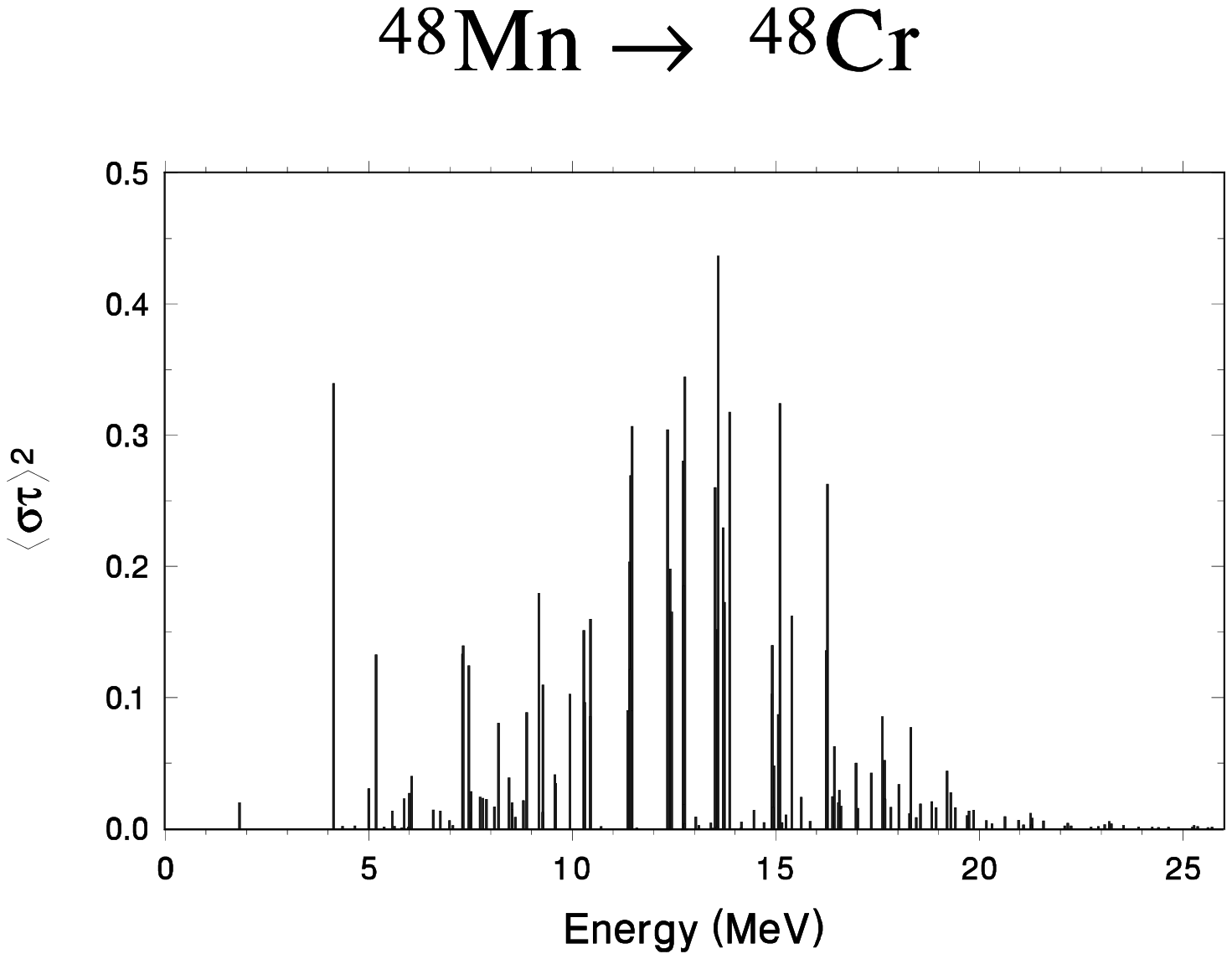}
  \caption{Same as figure 12 but without any binning}
  \label{fig:mndet}
\end{figure}

To cure the discrepancies we need a mechanism that may
affect very slightly the overall GT distribution, without affecting
the positions and other properties of the underlying levels, which are
very well reproduced. Here, it is useful to remind how the strength
function is obtained.

First, we define a state $|s\rangle$ by acting on the parent,
$|s\rangle = \sigma\tau|i\rangle S^{-1/2}$ which exhausts by
construction the sum rule $S=\langle i|(\sigma\tau)^2|i\rangle$, and
then evaluate the amplitude of $|s\rangle$ in each daughter state by
doing Lanzos interactions using $|s\rangle$ as pivot~\cite{cpz1}. The
first iteration produces the centroid and variance of the strength
function, $E_s=\langle s |H|s \rangle$ and $v_s=\langle
s|H^2-E_s^2|s\rangle$ respectively. As the number of iterations $\nu$
increases, the strength ---originally concentrated at $E_s$--- is
fragmented into $\nu$ peaks. The lowest converge to exact eigenstates
(with their exact share of the total sum rule) at the rate of roughly
one every 6-10 iterations.  Figure~\ref{fig:mndet} is the high
resolution version of fig.~\ref{fig:mncrt} and shows the situation
after 60 iterations for each of the $JT$ values ($J=3,4,5$;
$T=0,1,2$). It is only below approximately 6~MeV that we have a
complete picture of the spectrum.  Above, many thousand states are
waiting to come and erode the strong peaks.

Let us examine first the global properties associated with $S$ and
$E_s$ and then turn to the consequences of local fragmentation.

\begin{description}

\item[Quenching.] Calculations always produce too much strength, that
  has to be reduced by a quenching factor, which is the stronger
  (i.e.\ smaller) the most drastic the truncation~\cite{cpz1}. For
  exact calculations in {\em one major shell\/} ($0\hbar\omega$), $S$
  has to be reduced by $(0.77)^2$, {\em which is very much the value
    demanded by the ``violation'' of the model independent sum rule\/}
  (\ref{eq:sumrule}). There is very little we can do {\em within a
    $0\hbar\omega$ calculation\/} to change this state of affairs.

\item[Shifting.] Contrary to $S$, which depends on geometry
  (\ref{eq:sumrule}) and on overall properties of $H$, $E_s$ may be
  significantly affected by small changes in the $\sigma\cdot\sigma$
  and $\sigma\tau \cdot \sigma\tau$ contributions to $H$. In
  ref.~\cite{duzu} it is shown that these spin-spin terms are very
  strong, especially the second, and may differ from force to force by
  some 20\% (the corresponding constants $e_{1^+0}$ and $e_{1^+1}$ in
  ref.~\cite{duzu} are related, but not trivially, to the Migdal
  parameters $g$ and $g'$). Therefore, the mechanism to cure the small
  discrepancies we have mentioned may well come from modifications in
  these numbers, that would produce small overall {\em shifts\/} of
  the distribution, and nothing else.
\end{description}

Now we return to the quenching problem. In view of discrepancies
between $(p,n)$ and $\beta^+$ data for $^{37}$Ca, the extraction of
$S$ from the former has been recently criticized~\cite{adel,auf}. The
problem was compounded by the fact that calculations with Wildenthal's
W interaction suggested that standard quenching did not seem
necessary.  This illusion was dispelled by Brown's
analysis~\cite{brown}, showing that the interaction was at fault. It
is very interesting to note that what is called 12.5p in~\cite{brown}
is none other that KB, while CW ---which gives the best results--- is
very basically a minimally modified KB that can be safely assimilated
to KB1 or KB3.

\begin{description}
\item[Diluting.] Brown's analysis contains another important hint:
  once normalized by the $(0.77)^2$ factor, the CW calculations follow
  smoothly the data within the $\beta$ window, but then produce too
  much strength when compared with the $(p,n)$ reaction. The
  suggestion from fig.~\ref{fig:mndet} is that in regions where the
  density of levels becomes high, fragmentation may become so strong
  that much strength will be so {\em diluted\/} as to be rendered
  unobservable. The difference between the A~=~37 and A~=~48 spectra
  is that in the former the effect is almost exclusively due to
  intruders, while in the latter the density of $pf$ states is high
  enough to produce substantial dilution by itself. Which brings us
  back to the problem of the amount of strength calculated between
  8.75 and 11.75~MeV in $^{48}$Cr, double the measured value.  Some
  shifting may be warranted to reduce the discrepancy, but
  fig.~\ref{fig:mndet} suggest very strongly that dilution be made
  responsible for it (i.e.\ for anomalous quenching)
\end{description}

{}From all this, it follows that simultaneous measurements of $\beta^+$
and $(p,n)$ strength are very much welcome in pairs of conjugate
nuclei where the $\beta^+$ release energy is large, the $0\hbar\omega$
spectrum is dense and high quality calculations are feasible. We
propose the following candidates:

\begin{center}
\begin{tabular}[h]{cc}
$\beta^+$ & $(p,n)$ \\
$^{45}$Cr $\rightarrow$ $^{45}$V & $^{45}$Sc $\rightarrow$ $^{45}$Ti
\\
$^{46}$Cr $\rightarrow$ $^{46}$V & $^{46}$Ti $\rightarrow$ $^{46}$V
\\
$^{47}$Mn $\rightarrow$ $^{47}$Cr & $^{47}$Ti $\rightarrow$ $^{47}$V
\\
$^{48}$Fe $\rightarrow$ $^{48}$Mn & $^{48}$Ti $\rightarrow$ $^{48}$V.
\\
\end{tabular}
\end{center}

To conclude: a theoretical understanding of standard quenching demands
that we look at the full wavefunction and not only at its
$0\hbar\omega$ components. Experimentally, what has to be explained is
the disappearance of strength, i.e.\ standard {\em and\/} anomalous
quenching (as observed in $^{48}$Cr). Dilution will no doubt play a
role in both, but the latter may be observed already by comparisons
with $0\hbar\omega$ calculations.

\section{The validity of {\sl monopole} modified realistic
  interactions}

To answer with some care the question raised in the introduction we
review briefly the work related to monopole corrections.

The first attempt to transpose the results of refs.~\cite{pazu1,pazu2}
to the sd shell met with the problem that the interaction had to
evolve from $^{16}$O to $^{40}$Ca. A linear evolution was assumed, but
it was shown that the centroids followed more complicated laws
demanding an excessive number of parameters~\cite{corzu}.

The solution came in ref.~\cite{acz}, by adopting a hierarchy of
centroids and noting that the realistic matrix elements depend on the
harmonic oscillator frequency very much as

\begin{equation}
W_{rstu}^{JT} (\omega)=\frac{\omega}{\omega_0} W_{rstu}^{JT}(\omega_0),
\end{equation}
thus displacing the problem of evolution of $H$ to one of evolution of
$\omega$. The classical estimate $\hbar\omega=40A^{-1/3}$~\cite{bohr}
relies on filling oscillator orbits and on adopting the $r=r_0
A^{1/3}$ law for radii, which nuclei in the $p$ and $sd$ shells do not
follow. Therefore it was decided to treat $\omega$ as a free parameter
for each mass number and then check that the corresponding oscillator
orbits reproduce the observed radii. This turned out to work very well
and to produce very good spectroscopy in all regions where exact
calculations could be done.

The method relies on the rigorous decomposition of the full
Hamiltonian as ${\cal H}={\cal H}_m+{\cal H}_M$, where the monopole
part ${\cal H}_m$ is responsible for saturation properties, while
${\cal H}_M$ contains all the other multipoles.  Upon reduction to a
model $H,{\cal H}_m$ is represented by $H_m$, which contains the
binding energies of the closed shells, the single particle energies
and the centroids.  Everything else goes to $H_M$, which nevertheless
depends on ${\cal H}_m$ through the orbits, in principle
selfconsistently extracted from ${\cal H}_m$~\cite{acz,duzu,zuker}.
The program of minimal modifications now amounts to discard from the
nucleon-nucleon potential the ${\cal H}_m$ part and accept all the
rest, unless some irrefutable arguments show up for modifying
something else. On the contrary ${\cal H}_m$ is assumed to be purely
phenomenological and the information necessary to construct it comes
mostly from masses and single particle energies~\cite{zuker}.

The proof of the validity of the realistic ${\cal H}_M$ through shell model
calculations depends on the quality of the monopole corrections. In
regions of agitated radial behaviour we have to go beyond the
oscillator approximation. In particular, the sd shell radii can be
reproduced practically within error bars by Hartree-Fock calculations
with Skyrme forces with orbital fillings extracted from the shell
model wave functions~\cite{abzo}.  Obviously the $d_{5/2},s_{1/2}$ and
$d_{3/2}$ orbits are poorly reproduced by a single $\omega$, and
obviously this makes a difference in the two body matrix
elements~\cite{kls} (work is under progress on this problem).

Therefore, in the sd shell ---to match or better the energy agreements
obtained with Wildenthal's famous $W$ (or USD)
interaction~\cite{wilden,browil}--- we have to push a bit further the
work of ref.~\cite{acz}.

When we move to the $pf$ shell, no such efforts are, or were,
necessary.  Because of $f_{7/2}$ dominance, it is much easier to
determine the centroids: the $V_{rr}^T$ values are no issue, the
crucial $V_{fr}^{T}$ ones can be read (almost) directly from single
particle properties on $^{49}$Ca and $^{57}$Ni, and we are left with
$V_{ff}^T$; only two numbers. Once we have good enough approximations
for the centroids we can do shell model calculations to see how the
rest of $H$ (i.e. $H_M$) behaves.  In ref.~\cite{pozu2} it was found
that $H_M$ behaves quite well, but much better in the second half of
the $f^n$ region: A~=~48 happens to be the border beyond which quite
well becomes very well.  Hence the remark in the second paragraph of
the introduction.

The trouble at the beginning of the region was attributed at the time
to intruders, but now we know that radial behaviour must be granted
its share. It is here that KB3 comes in. Although eq.~\ref{eq:kb3} has
only cosmetic effects, its origin is not cosmetic: it was meant to
simulate necessary corrective action {\it not related to the presence
  of intruders}.  It reflects the fact that in the neighborhood of
$^{40}$Ca the interaction works better if the spectrum of $^{42}$Sc is
better reproduced, while at A~=~48 and after, the KB interaction
---not very good in $^{42}$Sc--- needs no help, except in the
centroids, as we have shown.

Most probably this says something about radial behaviour at the
beginning of the region. It is poorly known except in the Ca isotopes,
where it is highly non trivial~\cite{cpz4}. The indications are that
when shell model calculations demand individual variations of matrix
elements in going from nucleus to nucleus the most likely culprits are
the single particle orbits.  And the indications are that at
A~=~48 the need for such considerations disappears.

All this is to say that A~=~48 happens to be a good place for a test
of the interaction and KB1 passes it with honours. It also amounts to
say that it is not an accident but part of the general statement that
monopole corrected realistic interactions are the ones that {\em
  should\/} be used in calculations.

It is important to stress that monopole corrections are both necessary
and sufficient.

Doing less amounts to run great risks as illustrated by the $T=2$
states in A~=~56. Using KB leads to a very good looking spectrum in a
$t=2$ calculation of $^{56}$Fe~\cite{otsuka}, but ignores the fact
that the force would produce nonsense for $T=0$. (For $T=2$, KB1 or
KB3 gives results that are even better looking than those of KB, and
somewhat different).

Doing more, under the form of extended fits of all matrix elements,
has become unnecessary and may be misleading as illustrated by the
problems encountered by W, the most famous of the fitted interactions
(section 9,~\cite{brown}). Still, a full explanation of its success in
terms of monopole behaviour remains a challenge.

\section{Note on binding energies}

Our interaction overbinds all the A~=~48 nuclei by about the same
amount, indicating the need of small corrections in the centroids. The
addition of an overall $Vn(n-1)/2$ term with $V=28$~keV, leads to the
binding energies in table~\ref{tab:binding}, which have been Coulomb
corrected by subtracting $(n=\pi+\nu,\pi$ = protons, $\nu$ =
neutrons)~\cite{pazu1}

\begin{eqnarray}
H_{\mbox{\scriptsize coul}}&=&
V_{\pi\pi}\pi(\pi-1)/2+V_{\pi\nu}\pi\nu+7.279\pi
\mbox{ MeV} \nonumber \\
V_{\pi\pi}&=&0.300(50)\mbox{ MeV}, V_{\pi\nu}=-0.065(15)\mbox{ MeV.}
\end{eqnarray}

\begin{table}[h]
  \begin{center}
    \leavevmode
    \begin{tabular}{lccccccc}
      \hline \hline
      & Ca & Sc & Ti & V & Cr & Mn & Fe \\ \hline
      Exp & 7.04 & 13.35 & 23.82 & 26.70 & 32.38 & 26.86 & \\
      Th & 7.10 & 13.35 & 23.79 & 26.80 & 32.17 & 26.80 & 23.79 \\
      \hline \hline
    \end{tabular}
  \end{center}
  \caption{Nuclear two-body energies (MeV), A~=~48}
  \label{tab:binding}
\end{table}

For $^{48}$Fe we obtain $BE=385.37\,$MeV, leading to a decay energy of
10.94$\,$MeV for \linebreak $^{48}$Fe~$\rightarrow$~$^{48}$Mn.  The
most recent estimate of Wapstra, Audi and Hoekstra~\cite{wap} gives
\linebreak $BE(^{48}$Fe)=385.21~MeV. The precision of
subtractions~\cite{rich} can be checked through the position of the
analog of $^{48}$V in $^{48}$Cr, experimentally at 5.79$\,$MeV against
the Coulomb corrected value of 5.68$\,$MeV in table~\ref{tab:binding}.
As noted in section~8, for this state there is a 400$\,$keV
discrepancy between experiment and calculation, high by our standards.
It probably indicates the need of some more sophistication in the
readjustment of monopole terms than the 28$\,$keV we have suggested.
Note that the theory-experiment differences in table~\ref{tab:binding}
are typical of the results throughout the paper.

\section{Conclusions}

Since much of the sections 9 and 10 was devoted to drawing conclusions
about the two main problems addressed, we shall only sum them up:

\begin{itemize}
\item[i)] The detailed agreement with the data lends strong support to
  the claim that minimally monopole modified realistic forces are the
  natural and correct choice in structure calculations.
\item[ii)] The description of Gamow-Teller strength is quite
  consistent with the data, to within the standard quenching factor.
  The calculations strongly suggest that dilution due to fragmentation
  makes much of the strength unobservable.
\end{itemize}

Two by-products emerge from the calculations. The first is mainly
technical:

\begin{itemize}
\item[iii)] Truncations at the $t=3$ level are reasonable in the lower
  part of the $pf$ shell. In general, however, truncations are a
  dangerous tool and it would be preferable to replace them by some
  other approximation method.
\end{itemize}

The second by-product is more interesting:

\begin{itemize}
\item[iv)] The calculations provide  a very clean microscopic view of
  the notion of intrinsic states and of the conditions under which
  rotational motion sets in.
\end{itemize}

\subsection*{Acknowledgements}

This work has been supported in part by the DGICYT (Spain) grant
PB89-164 and by the IN2P3 (France)-CICYT (Spain) agreements.


\begin{thebibliography}{99}

\bibitem{gifr} J.N.~Ginocchio and J.B.~French,
 Phys. Lett.
 {\bf 7} (1963) 137.\\
 J.N.~Ginocchio,
 Phys. Rev.
 {\bf 144} (1966) 952.

\bibitem{mbz} J.D.~Mc~Cullen, B.F. Bayman and L.Zamick,
 Phys. Rev.
 {\bf B4} (1964) 515.

\bibitem{hoog} H.~Horie and K.~Ogawa,
 Nucl. Phys.
 {\bf A216} (1973) 407.

\bibitem{pazu1} E.~Pasquini,
Ph. D. Thesis CRN/PT 76-14, Strasburg 1976.

\bibitem{pazu2} E.~Pasquini and A.P.~Zuker,
``Physics of Medium Light Nuclei'',
Florence 1977,
 P.~Blasi and R.~Ricci eds. (Editrice compositrice,
Bologna 1978).

\bibitem{glau1} A.G.M.~van Hess and P.W.M.~Glaudemans,
 Z. Phys.
 {\bf A303} (1981) 267.

\bibitem{glau2} R.B.M.~Mooy and P.W.M.~Glaudemans,
 Z. Phys.
 {\bf A312} (1983) 59.

\bibitem{glau3} R.B.M.~Mooy and P.W.M.~Glaudemans,
 Nucl. Phys.
 {\bf A438} (1985) 461.

\bibitem{hori1} K.~Muto and H.~Horie,
 Phys. Lett.
 {\bf B138} (1984) 9.

\bibitem{hori2} A.~Yokohama and H.~Horie,
 Phys. Rev.
 {\bf C31} (1985) 1012.

\bibitem{hori3} K.~Muto,
 Nucl. Phys.
 {\bf A451} (1986) 481.

\bibitem{halb1} J.B.~Mc~Grory, B.H.~Wildenthal and E.C.~Halbert,
 Phys. Rev.
 {\bf C2} (1970) 186.

\bibitem{halb2} J.B.~Mc~Grory and B.H.~Wildenthal,
 Phys. Lett.
 {\bf B103} (1981) 173.

\bibitem{mcgr} J.B.~Mc~Grory,
 Phys. Rev.
 {\bf C8} (1973) 693.

\bibitem{cole} B.J.~Cole,
 J. Phys.
 {\bf G11} (1985) 481.

\bibitem{rich} W.A.~Richter, M.G.~Van der Merwe, R.E.~Julies and
B.A.~Brown,
 Nucl. Phys.
 {\bf A523} (1991) 325.

\bibitem{cpz1} E.~Caurier, A.~Poves and A.P.~Zuker,
 Phys. Lett.
 {\bf B256} (1991) 301.

\bibitem{ogho} K.~Ogawa and H.~Horie,
``Nuclear weak processes an nuclear structure'',
 M.~Morita et al.\ eds. (World Scientific,
Singapore 1990).

\bibitem{cpz2} E.~Caurier, A.~Poves and A.P.~Zuker,
 Phys. Lett.
 {\bf B252} (1990) 13.

\bibitem{enge} J.~Engel, W.C.~Haxton and P.~Vogel,
 Phys. Rev.
 {\bf C46} (1992) 2153.

\bibitem{kubro} W.~Kutschera, B.A.~Brown and K.~Ogawa,
 Riv. Nuovo Cimento Vol.
 {\bf 1} n$^{\rm o}$ 12 (1978).

\bibitem{pozu1} A.~Poves, E.~Pasquini and A.P.~Zuker,
 Phys. Lett.
 {\bf B82} (1979) 319.

\bibitem{pozu2} A.~Poves and A.P.~Zuker,
 Phys. Rep.
 {\bf 70} (1981) 235.

\bibitem{burrows} T.W.~Burrows,
 Nuclear Data Sheets
 {\bf 68} (1993) 1.

\bibitem{antoine} E.~Caurier,
Code ANTOINE
(Strasburg, 1989).

\bibitem{cpz3} E.~Caurier, A.~Poves and A.P.~Zuker,
 Proc.\ of the Workshop ``Nuclear Structure of Light Nuclei far from
 Stability. Experiment and Theory.''
 Obernai.
 G.~Klotz ed.
(CRN, Strasburg, 1989).

\bibitem{white} R.R.~Whitehead
in ``Moment methods in many fermion systems''
 B.J.~Dalton et al.\
eds. (Plenum, New York, 1980).

\bibitem{kuob} T.T.S.~Kuo and G.E.~Brown,
 Nucl. Phys.
 {\bf A114} (1968) 241.

\bibitem{edmonds} A.R.~Edmonds,
``Angular Momentum in Quantum Mechanics''
(Princeton, N.J. 1960).

\bibitem{wilki} D.H.~Wilkinson and B.E.F.~Macefield,
 Nucl. Phys.
 {\bf A232} (1974) 58.

\bibitem{bamby} W.~Bambynek, H.~Behrens, M.H.~Chen, B.~Graseman,
M.L.~Fitzpatrick,
K.W.D.~Ledingham, H.~Genz, M.~Mutterer and R.L.~Intemann
 Rev. Mod. Phys.
 {\bf 49} (1977) 77.

\bibitem{flem} D.G. Fleming, O. Nathan, H.B. Jensen and O. Hansen,
Phys. Rev. {\bf C5} (1972) 1365.

\bibitem{radga} P.~Raghavan,
 Atomic Data and Nuclear Data Tables
 {\bf 42} (1989) 189.

\bibitem{kls} S.~Kahana, H.C.~Lee and C.K.~Scott,
 Phys. Rev.
 {\bf 180} (1969) 956.

\bibitem{acz} A.~Abzouzi, E.~Caurier and A.P.~Zuker,
 Phys. Rev. Lett.
 {\bf 66} (1991) 1134.

\bibitem{duzu} M.~Dufour and A.P.~Zuker,
preprint CRN 93-29, proposed to Physics Reports, (1993).

\bibitem{lee} H.C.~Lee.
{\it private communication}, (1969).

\bibitem{roeck1} T.~Sekine, J.~Cerny, R.~Kirchner, O.~Klepper,
V.T.~Koslowsky, A.~Plochocki, E.~Roeckl, D.~Schardt, B.~Shenill and
B.A.~Brown,
Nucl. Phys.  {\bf A467} (1987) 93.

\bibitem{roeck2} J.~Szerypo, D.~Bazin, B.A.~Brown,
D.~Guillemaud-Muller, H.~Keller, R.~Kirchner, O.~Klepper,
D.~Morrissey, E.~Roeckl, D.~Schardt, B.~Sherrill, Nucl. Phys.  {\bf
  A528} (1991) 203.

\bibitem{adel} E.G.~Adelberger, A.~Garcia, P.V.~Magnus and D.P.~Wells,
 Phys. Rev. Lett.
 {\bf 67} (1991) 3658.

\bibitem{auf} M.B.~Aufderheide, S.D.~Bloom, D.A.~Resler and
D.C.~Goodman,
 Phys. Rev.
 {\bf C46} (1992) 2251.

\bibitem{brown} B.A.~Brown, Phys. Rev. Lett. {\bf 69} (1992) 1034.

\bibitem{corzu} A.~Cort\'{e}s and A.P.~Zuker,
 Phys. Lett.
 {\bf B84} (1979) 25.

\bibitem{bohr} A.~Bohr and B.~Mottelson,
``Nuclear Structure''
(Benjamin, NY 1969).

\bibitem{zuker} A.P.~Zuker,
preprint CRN 92-16, to be published in
 Nucl. Phys.

\bibitem{abzo} A.~Abzouzi, J.M.~Gomez, C.~Prieto and A.P.~Zuker,
CRN Strasbourg Annual Report (1990) p.15.

\bibitem{wilden} B.H.~Wildenthal,
 Prog. Part. Nucl. Phys.
 {\bf 11} (1984) 5.

\bibitem{browil} B.A. Brown and B.H.~Wildenthal,
 Annual Rev. Nucl.\ and Part. Sci.
 {\bf 38} (1988) 292.

\bibitem{cpz4} E.~Caurier, A.~Poves and A.P.~Zuker,
 Phys. Lett.
 {\bf B96} (1980) 15.

\bibitem{otsuka} H.~Nakada, T.~Otsuka and T.~Sebe, Phys. Rev. Lett.
{\bf 67} (1991) 1086.

\bibitem{wap} A.H.~Wapstra, G.~Audi and R.~Hoekstra,
 Atomic Data and Nuclear Data Tables
 {\bf 39} (1988) 281.


\end{thebibliography}
\end{document}